\RequirePackage{fix-cm}
\documentclass[twocolumn]{svjour3}          
\smartqed  

\usepackage{mathptmx}      		
\usepackage {etoolbox}			
\usepackage[misc,geometry]{ifsym} 	
\usepackage[numbers]{natbib}		
\usepackage{amsfonts}
\usepackage{amsmath}			
\usepackage{diagbox}
\usepackage{subfig}				
\usepackage{array}				
\usepackage{graphicx}
\usepackage[title]{appendix}		

\newcommand*{\affaddr}[1]{#1} 
\newcommand*{\affmark}[1][*]{\textsuperscript{#1}}

\begin{document}

\title{A Trust Management Scheme for IoT-Enabled Environmental Health/Accessibility Monitoring Services}

\author{Behshid Shayesteh\affmark [1] \and Vesal Hakami\affmark [1] \Letter \and Ahmad Akbari\affmark [1]}

\institute{Behshid Shayesteh\\
              \email{behshid.shayesteh@gmail.com}           
           \\
           Vesal Hakami \\
              \email{vhakami@iust.ac.ir}
              \\
              Ahmad Akbari \\
              \email{akbari@iust.ac.ir}
              \\
              \affaddr{\affmark[1] School of Computer Engineering, Iran University of Science and Technology, Tehran, Iran}
}

\date{Received: date / Accepted: date}

\maketitle

\begin{abstract}
One rapidly growing application of Internet of \linebreak Things (IoT) is the protection of public health and well-being through enabling environmental monitoring services. In particular, an IoT-enabled health/accessibility monitoring service (HAMS) can be consulted by its users to query about the status of different areas so as to optimize their trip throughout a geographic region. Given the high cost associated with a vast deployment of totally trusted information sources, the IoT-enabled monitoring services also subsist on citizen engagement and on (possibly untrusted) users’ sensing apparatus for data collection. However, trust management becomes a key factor in the success of such services because they might be misled by malicious users through altered or fake sensor data. In this paper, we consider a monitoring service, and propose a hybrid entity/data trust computation scheme which relies on Bayesian learning to score the users (as data reporters), and on Dempster-Shafer theory (DST) for data fusion and for the computation of the trustworthiness of the data itself. In order to provide resiliency against behavioral changes, the probability masses used in DST are dynamically updated using the freshly estimated user scores as well as the contextual properties associated with the reported data. We conduct simulation experiments to evaluate the performance of our scheme. Compared to prior work, the results demonstrate superior performance in terms of accuracy and resilience against malicious behavior
\end{abstract}


\section{Introduction}
\label{sec:1}

\subsection{Research Background}
\label{sec:1.1}
The origination and aggravation of health problems due to environmental exposures are complex and multifactorial. Numerous environmental exposures have been identified as contributing factors, including food and water safety, ambient levels of respirable particulate matter, hazardous chemicals such as pesticides and allergens, etc. \cite{dieffenderfer2016low}. An increased attention needs to be paid to the complex environmental issues that affect the health and well-being of people, and recent research suggests greater civic engagement in processes that shape public health. In fact, a successful health/accessibility monitoring service (HAMS) should be a partnership approach that involves, for example, community members, organizational representatives, and clinical social workers. These services are becoming increasingly common thanks to the rapidly growing deployment of Internet of Things (IoT) along with a synergistic adoption of smart sensing and cloud computing paradigms \cite{borgia2014internet} \cite{al2015internet}.

\par In an IoT-enabled HAMS, there is a cloud-computing-based remote aggregator in which the data emanated from multiple device sensors and/or human observers, covering one or more than one place, are collated, analyzed, contextualized (including with other non-sensor data) and summarized as appropriate. This central server can then be queried to provide information and to support decision-making tasks of various kinds. Consider the parents of a child with asthma, they could consult the monitoring service to obtain knowledge of places where the concentration of respiratory irritants is high. Similarly, the service can be used to dynamically discover, assess and classify urban accessibility issues, generating valuable information to improve citizens’ quality of life. Examples are the presence/absence of an accessibility barrier/facility (e.g., steps and stairs that can be detected by walking pedestrians, ramps and curb cuts that can be detected by wheelchair users, etc.).

\par Owing to their inclusive nature, IoT-enabled monitoring systems opt for contributions from people of all backgrounds, effectively from anyone possessing a \linebreak sensing-capable device. However, the volume, potential for anonymity, and lack of context can make the information derived from citizen participation hard to trust, and its intent and origin hard to discover. The lack of control mechanisms to guarantee source validity and data accuracy can result in information credibility issues. For example, malicious individuals may intentionally contribute erroneous sensing data for their own benefit. Therefore,  it  is  necessary  to  develop  trust preservation  and  abnormality  detection  technologies  to  ensure the quality of the obtained data. Realizing this requirement, in this paper, we study the issue of trust management for IoT-enabled environmental monitoring services. Before stating our contributions, we first give an overview of the state of the art on trust management in IoT to highlight their strengths and shortcomings.

\vfill

\subsection{Related Works}
\label{sec:1.2}
Trustworthiness in IoT-enabled services arises as a crucial concern because these services might be misled by malicious users through altered or fake sensor data. Nonetheless, coming up with an effective trust management scheme for IoT applications is challenging primarily due to the issue of data inconsistency; in fact, it is quite common that reliable sensors, faulty or compromised sensors generate distinct readings for the same observed phenomenon. As a result, the truth must be inferred from fusing (contradictory) data, originating from untrustworthy sources describing dynamic and uncertain phenomena for which we may not have a fine-tuned prior statistical description or model. Even more challenging is the need to assess and sift faulty data without any assumption on the trustworthiness of their source. In what follows, we survey the research works that have addressed the issue of trust in IoT-enabled services.

\par In one broad taxonomy, we may categorize the research on IoT trust management into three areas: \textit{entity-centric}, \textit{data-centric}, and \textit{hybrid} schemes. We elaborate on each category with references to some representative research in general as well as in the context of health-IoT and urban crowd-sensing.

\subsubsection{Entity-centric Trust Management}
\label{sec:1.2.1}
The entity-centric trust management focuses on assessing the trust in the participants of the IoT system considering their behavioral tendencies. The purpose of computing entity-centric trust is to identify the malicious or selfish users, which may tend to initiate an attack or compromise the service being provided. In \cite{guo2014classification}, a survey is conducted on the entity-centric trust management, and the authors provide a classification of trust computation techniques in IoT. In fact, we may further sub-classify the entity-centric schemes depending on whether trust computation is performed in a distributed fashion or otherwise by a centralized authority:

\par In distributed trust computation, it is most common that the social relationships between entities are taken into account. For example, the study in \cite{truong2016reputation} provides a trust service platform for the social IoT, which uses recommendation and reputation trust metrics, as well as knowledge trust metrics. They estimate trust scores for the entities in their system, using a fuzzy-based approach. Bao and Chen in \cite{bao2012dynamic} propose a dynamic trust management to deal with misbehaving nodes whose behavioral tendencies change dynamically in an IoT environment, and adjust the trust parameters adaptively using a static weighted sum approach. Similarly, Chen et al. provide an adaptive trust management system based on the common community of interests for social IoT in \cite{bao2013scalable}, which uses a dynamic weighted sum method to assess the trust scores. The work in \cite{nitti2013trustworthiness} provides one subjective and one objective trustworthiness management model for social IoT using the static weighted sum approach. In subjective trustworthiness model, users or entities compute the trustworthiness of their friends; however, in objective trustworthiness model, the trust is computed globally. Chen et al. propose an adaptive trust management system to support service composition application in SOA-based IoT systems in \cite{chen2016trust}. They compute trust value for service providers, which can be the user’s devices, based on experience and recommendations to offer a trust-based service composition. They adopt a Bayesian framework for evaluating direct trust, and a dynamic weighted sum for indirect trust.

\par Centralized trust computation, on the other hand, is motivated by the high costs associated with a fully distributed scheme in terms of processing power, memory, bandwidth, and energy consumption given that each node not only has to monitor the behaviors of other nodes but also manage the trust records for them, which do not meet the resource constraints of the network. In \cite{saied2013trust}, a trust management scheme is proposed for service-oriented IoT using a dynamic \linebreak weighted sum approach, in which a node is provided with the best service assistant node in a specific context. A recommendation trust score is calculated for the recommendation provider entity, based on the deviation of this report from the majority. Also in \cite{he2012retrust}, a medical sensor network has been considered for which an attack-resistant lightweight trust management protocol has been proposed. Each node submits all its trust records to a base station which runs an alerts reasoning algorithm to detect the malicious nodes using a weighted sum method.

\subsubsection{Data-centric Trust Management}
\label{sec:1.2.2}
The data-centric trust management focuses on assessing the trustworthiness of data or events in IoT and on detecting erroneous data. Since data analysis is an important phase in IoT solutions, it is essential to make sure that the data used in the decision making process, or in the actuators, are trustworthy and reliable. However, compared to entity-centric trust management, fewer works have approached IoT trust computation purely from a data-centric perspective.

\par In \cite{liu2015toward}, an IoT agriculture scenario has been considered and the authors devise a procedure to distinguish reliable sensor data from unreliable data to be further used in data analysis. In this scenario, the temperature and humidity data are gathered from the sensors deployed in a greenhouse to a central unit. It is argued that the sensors may gradually become unreliable due to the changes in the environment, and thus the authors have proposed to compute the trustworthiness of data using a dynamic Bayesian approach. However, the work in \cite{liu2015toward} does not consider temporal or spatial context parameters for assessing data trustworthiness, and therefore, it is only applicable to a subset of IoT scenarios with no user participation. In \cite{prandi2017need}, a data trustworthiness assessment method has been proposed for a system that collects data from crowdsourcing and sensing to map urban and architectural accessibility. This work evaluates the trustworthiness of each report about a barrier/facility based on the contextual parameters of the reporter using a static weighted sum approach.

\subsubsection{Hybrid Trust Management}
\label{sec:1.2.3}
As the IoT environment consists of entities interacting with the services, and also abundant data is generated and used for the decision making process, a trust management system would be most effective if it computes trust values for both the entities and the data. In hybrid trust management, the entity trust is maintained over time and it will be utilized as one important factor in determining data trustworthiness. However, there are other aspects such as timeliness, locality, and other contextual properties unique to data which also affect data quality, and thus need to be factored into the calculation of data trustworthiness.

\par The work in \cite{jayasinghe2017data} extends the entity-centric trust scheme in \cite{truong2016reputation} by proposing a framework in which data trust metrics such as: completeness, uniqueness, timeliness, accuracy, etc. are used to assess the data trustworthiness in a social IoT environment using a dynamic weighted sum method. However, the paper does not elaborate on the specifics of some critical parameters for trust calculation, and also lacks experimental evaluations. In \cite{li2018policy}, a policy-based hybrid trust management model called RealAlert has been proposed for smart city scenarios in IoT. The authors use a statistical outlier detection approach to identify malicious nodes and abnormal data in a distributed manner. They also devise some policies to find the cause of the abnormalities. The Dempster-Shafer Theory (DST)  \cite{shafer1976mathematical} has been applied to aggregate the abnormality reports that each node receives from it is neighbor; however, the paper is not clear on how to derive the basic probability masses for its DST-based data fusion. The authors in \cite{mahmud2018brain} introduce a neuro-fuzzy based brain-inspired trust management model to secure IoT devices and relay nodes, and to ensure reliable data communication between devices. The method in \cite{mahmud2018brain} evaluates both node behavioral trust (entity trust) and data trust. The entity trusts are computed in a distributed manner, and the social relationships between IoT devices are taken into account. However, unlike our proposed scheme in this paper, the method in \cite{mahmud2018brain} focuses on the brain data and neuroscience-related applications.

\par A more closely related work to our study is \cite{al2017trust} which provides a trust-based decision making for health-IoT solutions. In this work, users in an IoT environment contribute reports about health factors in an area. The central authority evaluates the trust of reporting users (i.e., entity trust), and spots the malicious users who may contribute false reports. The users may also query the central authority prior to entering an area to prevent worsening their health condition. The central authority then answers each query considering the data trustworthiness of the gathered reports to comprehend the healthiness of an area, and the health classification of each user, to minimize the risks for users’ health. While \cite{al2017trust} proposes a hybrid trust management for health-IoT applications, the authors use a simplistic method for computing entity trust scores. The basic idea is to compare a given node’s feedback with the majority of feedbacks about the same phenomenon in a location. Also, they have not evaluated their scheme in the presence of high percentage of malicious nodes in the system. It can be argued, however, that as more and more users exhibit unreliable behavior, the majority-based rationale is bound to become worse. Moreover, some parameters used for trust estimation in \cite{al2017trust} are not clearly specified (e.g., the parameters dealing with the temporal freshness of data and device capabilities).

\subsection{Motivations and Contributions}
\label{sec:1.3}
Major prior works on trust management in IoT rely on the notion of entity-centric trust, trivially taking the trust level of the data to be the same as the trust level of the data source. To the best of our knowledge, the only hybrid trust management scheme for health-IoT applications is \cite{al2017trust}; however, despite being a comprehensive framework, there is still room for enhancements to obtain more accurate trust estimates, especially in the presence of higher percentages of malicious users. In this paper, we propose a novel trust management scheme to run as a subsystem of a health/accessibility monitoring service (HAMS). Similarly to \cite{al2017trust}, we consider the case where there is a centralized server collecting observational reports on different environmental phenomena   from individual IoT devices. The users can also query the system about the health/accessibility status of different areas to protect their health/well-being while traversing the environment. Despite the similarity in use case scenario, our trust management scheme differs from \cite{al2017trust} when it comes to the computation of the entity trust scores as well as the data trustworthiness. In particular, we come up with the following contributions: 

\begin{itemize}
\item [$\bullet$] In contrast to the trivial majority-based estimate in \cite{al2017trust}, we propose a Bayesian learning approach for scoring the system entities. In particular, a given entity’s trust score is estimated according to its history of contributing behavior, gradually increasing the score for well-behaving nodes towards the high value, and thwarting misbehaving entities by lowering their value, ensuring the soft exclusion of their observations from the reasoning procedure. Also, unlike \cite{truong2016reputation,bao2012dynamic,bao2013scalable,nitti2013trustworthiness,chen2016trust} we propose a centralized scheme for estimating the entity trusts that offloads the overhead from the resource-constrained IoT devices to a central unit (possibly running in the cloud), and avoids much of the communication overhead.
\item [$\bullet$] As for the computation of data trustworthiness, we resort to the DST formalism for data fusion and to suppress malicious contributions and/or unintentionally noisy sensing data. The probability masses used in the DST’s rule of combination are derived from the freshly estimated entity trust scores as well as the contextual properties associated with the reported data. Compared to \cite{al2017trust}, where the aggregated trust is simplistically computed by a \linebreak weighted average procedure, DST is equipped by design to handle uncertainty or lack of complete information. Also, besides the usual contextual elements such as time and location (which have also been considered in \cite{al2017trust}), as an additional element, our work incorporates the semantic correlation between the groups of users and the phenomenon under observation.
\item [$\bullet$] We conduct extensive simulation experiments to explore the convergence behavior, the accuracy, and the resiliency of the proposed scheme, and to compare its performance with prior work. 
\end{itemize}

\subsection{Outline}
\label{sec:1.4}
The rest of the paper is organized as follows: In Sect.~\ref{sec:2}, we describe the system model, remark on some typical use case scenarios, and then lay out the main assumptions underpinning our scheme. In Sect.~\ref{sec:3}, we present our hybrid trust management subsystem for adoption in a HAMS. In particular, we first present our proposed scheme for the dynamic computation of entity trust scores, and then discuss how these scores, along with the contextual properties associated with each reported observation, can be incorporated into assessing data trustworthiness. Sect.~\ref{sec:3} ends with our proposed DST-based scheme for the computation of data trustworthiness. In Sect.~\ref{sec:4}, we present the simulation results along with comparisons that have been made against related prior work. Sect.~\ref{sec:5} concludes the paper.


\section{System Model and Assumptions}
\label{sec:2}

\subsection{Overview and Use-case Scenario}
\label{sec:2.1}
\begin{figure}
\includegraphics*[width=0.5\textwidth]{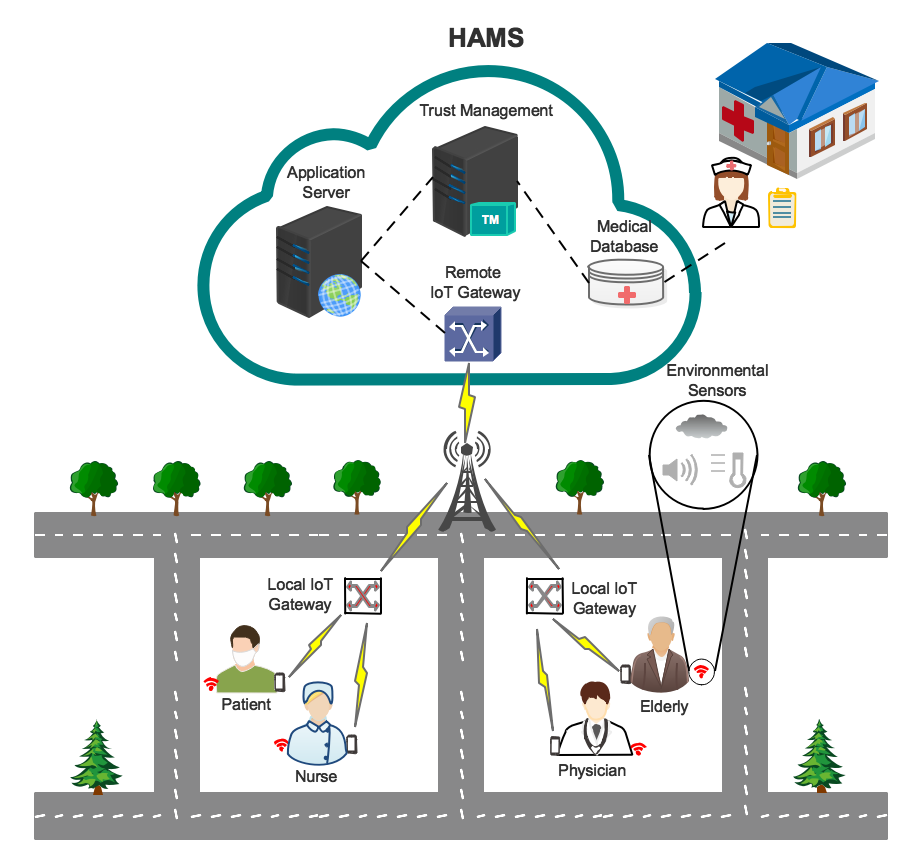}
\caption{IoT System Model}
\label{fig:1} 
\end{figure}

Fig.~\ref{fig:1} illustrates our IoT system model. It is assumed that the IoT environment is divided into $L$ different areas $A=\{a_1,\dots ,a_L\}$ based on their geographical coordinates. A total of $N$ users $U=\{u_1,\dots ,u_N\}$ exist in the environment who can generate observations about a set of environmental health/accessibility factors $F=\{f_1,\dots ,f_K\}$. Besides direct human-triggered input, there may be specialized sensor devices under the control of the users such as: environmental and weather sensors, either fixed (e.g., at home, in buildings, rooftops, etc.) or mobile (e.g., vehicle-mounted or held/worn by commuting user). It is further assumed that the environmental health/accessibility factors can be categorized into $Q$ different classes $C_F=\{c_1,\dots ,c_Q\}$ such as the factors related with urban health (e.g., the concentration of air pollutants, temperature, humidity, noise pollution, radiation etc.) or those impacting urban accessibility (e.g., facility maturity for disabled, sidewalk condition like narrowness or presence of rest areas, etc.). As for the users, we are mainly concerned with the following three assumptions:

\begin{itemize}
\item [$\bullet$] Each user $u_n$ owns an up-to-date health profile and the system has access to the user's latest records in hospitals, cancer, and clinical registries to subsequently link health status data with environmental risks. This database will provide all relevant information from the very basics (e.g., the user's age, gender, etc.) to more specific information like whether the user suffers from any special kind of disorder/disease. As also envisaged in \cite{al2017trust}, the medical database can be maintained with real-time information on the users' health level by way of recent technologies such as wearable sensors and using personal (body) area networks.
\item [$\bullet$] Users are categorized into $P$ different classes \linebreak $C_U=\{c_1,\dots ,c_P\}$. Such classification can be done according to different criteria; for instance, if a user is a healthcare professional, the most relevant criterion is \linebreak his/her authority level or scope of practice (e.g., physician, EMT, paramedic, etc.). Similarly, there can be city health officials (CHOs) tasked specifically with public health monitoring. In case of regular people, where no official classification applies, other relevant user classes can be defined (e.g., patients, vulnerable-groups, elderly, etc.).
\item [$\bullet$] Each user is associated with one or more home areas, which are the areas that the user frequently visits; e.g., the user's residential area or the user's working place. More formally, we use the function $home:U\to \ A$ as a mapping between a given user and its home area.
\end{itemize}

\par After user $u_n$ enters area $a_l$, the user makes an observation about the environmental factor $f_K$. It is further assumed that the users own smart phones which can be used as a personal gateway to transmit the observation data to a central unit called Health/Accessibility Monitoring Service (HAMS) through a local IoT gateway. By receiving different observations from different users about the environmental factors of a specific area, HAMS can reason about the health rate of that area. In fact, such reasoning can be subsequently exploited to reduce personal health risks, especially to the benefit of patients or other vulnerable classes of users; for instance, a user who is suffering from a chronic cardiopulmonary disease like asthma can consult HAMS and check the status of environmental health of some area to decide whether or not to enter that area. 

\par \textbf{Remark 1.} \textit{HAMS could assist a user in making a decision about entering an area, considering the user's health level and the health rate of the area, aiming to reduce the risk of having an attack or worsening user's health condition. As discussed in \cite{al2017trust}, a simple threshold process (such as the decision plane concept \cite{josang2004analysing}) could be used by the system to relate the risk factors with the users' health status, and to assess the probability of health loss in each case. However, in this paper, we mainly focus on the issues of trust management and are not concerned with the specifics of any particular decision support system.}

\subsection{User-System Interactions and Timings}
\label{sec:2.2}

\begin{figure*}[tbh]
\includegraphics[width=0.95\textwidth]{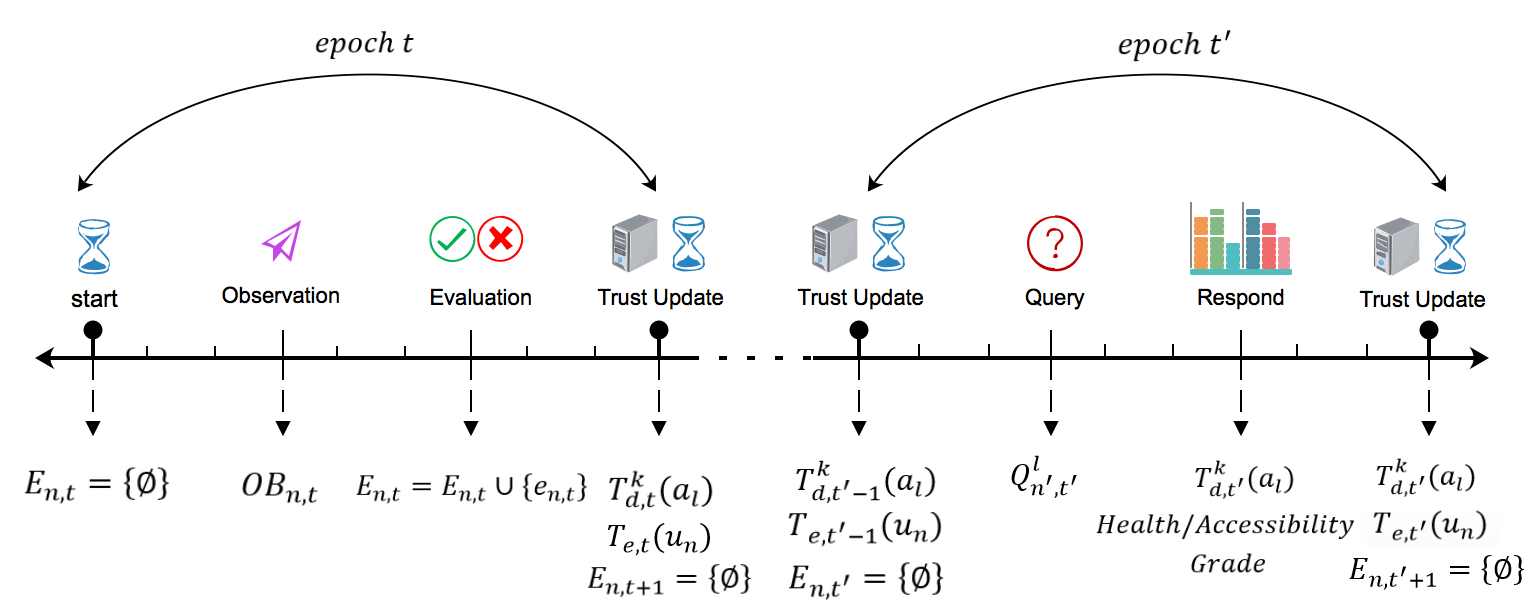}
\caption{Interaction Sequence of a User and HAMS}
\label{fig:2} 
\centering
\end{figure*}

Besides recruiting users, HAMS must also retain them by offering accurate and useful information. In fact, at any instant in our system model, a user can take on either of the two roles: to contribute observations or just query for the health/accessibility rate of an area in an on-demand fashion. We consider the case where besides honest users, there may also be several malicious/compromised users in the system who either deliberately contribute wrong observations or have faulty sensors that provide incorrect readings. We assess the trust value of each user to distinguish good users from the malicious. We assume that HAMS updates these trust values periodically in discrete epochs with length $t_{update}$ and indexed by $t=1,2,3,\dots $. This trust value is called entity trust as users are end entities in our system and is denoted by $T_{e,t}(u_n)$. Furthermore, it is also necessary to assess data trustworthiness of HAMS responses to users' queries as HAMS aggregates correct and incorrect observations about the factors of an area to infer the environmental health of that area. As such, if HAMS states that area $a_l$ is not healthy considering factor $f_k$, a data trustworthiness indicator $T^k_{d,t}(a_l)$ denotes the trustworthiness of this claim. The trust management module placed in HAMS is in charge of computations of both entity trust and data trustworthiness.

\par To better elucidate the user-system interactions, consider a sample epoch $t.$ We assume that a valid contributing behavior is for each user $u_n$ to send a total of $M^{l,k}$ observations on factor $f_k$ in area $a_l$ during each epoch $t$. Following the receipt of ${OB}^{l,k,i}_{n,t}$, i.e., the $i$-th observation of user $u_n$ about factor $f_k$ in area $a_l$, HAMS evaluates this observation by making a binary decision about its authenticity. More formally, we use $b^{l,k,i}_{n,t}$ to denote the reporting behavior of user $u_n$. Now, $b^{l,k,i}_{n,t}\mathrm{=}\mathbb{C}$ indicates that the user has contributed correct observations on factor $f_k$ in area $a_l$, and  $b^{l,k,i}_{n,t}=\mathbb{W}$ is used to indicate the action of sharing wrong observations. We use $E^{l,k}_{n,t}={\{e^{l,k,i}_{n,t}\}}_{i=1:M^{l,k}}$ to collect the set of binary evaluations that HAMS has made about all these observations shared by user $n$. In other terms, if $e^{l,k,i}_{n,t}=\ \mathbb{C}$ , it means that HAMS decides that user $u_n$ has shared correct information, and conversely for $e^{l,k,i}_{n,t}=\ \mathbb{W}$. 

\par It can be safely assumed that a HAMS relies partially on limited but totally trusted information sources (e.g., traffic cameras, environmental sensors deployed and managed by the authorities, etc.). These sources can offer accurate information but at a high deployment cost and with limited spatial coverage. Also, there may be some prior statistical description or model of the environmental factors. Overall, while HAMS can evaluate the received observations, its evaluations are prone to detection error and might deviate from the truth. In order to capture the possibility of misdetections, we introduce $f_P$ and $f_N$ to denote the probability of false positives (i.e., to mistakenly decide that a user has shared false observations) and false negatives (i.e., to mistakenly decide that a user has shared true observations) in HAMS's evaluations, respectively. Accordingly, the value of  $e^{l,k,i}_{n,t}$ can be expressed probabilistically as follows:

\begin{equation}
\label{eq:1}
\mathbb{P}\left[e^{l,k,i}_{n,t}|b^{l,k,i}_{n,t}\mathrm{=}\mathbb{C}\right]=
\begin{cases}
f_P, & \ e^{l,k,i}_{n,t}=\mathbb{W} \\
1-f_p, & \ e^{l,k,i}_{n,t}=\mathbb{C}
 \end{cases}
\end{equation}

\begin{equation}
\label{eq:2}
\mathbb{P}\left[e^{l,k,i}_{n,t}|b^{l,k,i}_{n,t}\mathrm{=}\mathbb{W}\right]=
\begin{cases}
f_N, & \ e^{l,k,i}_{n,t}=\mathbb{C} \\
1-f_N, & \ e^{l,k,i}_{n,t}=\mathbb{W}
 \end{cases}
\end{equation}

\par \textbf{Assumption 1.} \textit{For ease of expositions, it is assumed (without loss of generality) that for all users }$u_n$\textit{, the reporting behavior }$b_n$\textit{ is characteristic of }$u_n$\textit{ alone, and is independent of the area or the specific factor the user }$u_n$\textit{ is reporting on. Similarly, It is assumed that the parameters }$f_P$\textit{ and }$f_N$\textit{ are identical across all areas }$a_l\in A$\textit{ as well as all health/accessibility-related factors }$f_k\in F$\textit{. As such, for notational convenience, we may henceforth drop the superscripts from the evaluation set }$E^{l,k}_{n,t}$\textit{ and the reporting behavior }$b^{l,k}_{n,t}$\textit{.}

\par \textbf{Remark 2.} \textit{The parameters }$f_P$\textit{ and }$f_N$\textit{ can be obtained using training data and by way of a standard statistical parameter estimation method such as a Hidden Markov Model (HMM) process \cite{rabiner1986introduction}. More specifically, since the true user behavior }$b_{n,t}$\textit{ is never observable, the system can exploit the observed process }${OB}_{n,t}$\textit{ to estimate the hidden behavior sequence, either by finding the most likely one, or alternatively by using a-posteriori distributions over states. However, the specifics of such estimation are beyond the scope of this paper as we are mainly concerned with the higher level computations of entity trust and data trustworthiness. }

\par  The left half of Fig.~\ref{fig:2} depicts an example observation processing during an epoch. At the beginning of the $t$-th epoch, $E_{n,t}=\{\emptyset \}$. A user $u_n$ within area $a_l$, reports ${OB}^{l,k,i}_{n,t}$ which is the user's $i$-th observation about the $k$-th factor. HAMS then updates the set $E_{n,t}$ with the evaluation of this observation. On the other hand, the right half of Fig.~\ref{fig:2} shows how HAMS responds to an incoming query within another epoch $t'$:  Consider another user ${u_{n'}}$ who issues a query ${Q^l_{n',t'}}$ regarding the overall health status of area $a_l$. HAMS responds to this query based on the data trustworthiness indicator ${T^k_{d,t'}(a_l)}$. 

\par At the end of each epoch $t$, $T^k_{d,t}(a_l)$ and $T_{e,t}(u_n)$ will get updated, and the system proceeds to the next interval. Trust updates are carried out according to a computation procedure which we discuss in Sect.~\ref{sec:3}. 


\section{The Proposed Trust Management System}
\label{sec:3}
In this section, we introduce our proposed trust management (TM) subsystem to be employed in HAMS. As explained in the Introduction, we opt for a hybrid approach to trust management, where trust computations are carried out for both the system entities (as data observers) as well as for the values of each environmental factor (as aggregation of data observations). In hybrid trust management, the entity trusts need to be maintained over time, while at the same time they will be dynamically integrated into the evaluation of data trustworthiness. Our proposed TM subsystem operates in a central manner, i.e., all trust computations are performed by HAMS (e.g., using cloud computing), which relieves the end-users from extra computational burdens. 
\par Our presentation of the TM subsystem in HAMS is organized as follows: In sect.~\ref{sec:3.1}, we introduce a notion of ``scores'' to capture the users' contributing behavior, and \linebreak model the entity trust as the user's score. We then propose a Bayesian learning-based scheme for the dynamic computation of these scores over time. Next, in sect.~\ref{sec:3.2}, we introduce our proposed scheme for computing data trustworthiness. In particular, we discuss how to incorporate the entity trusts as well as the contextual properties associated with each received observation to obtain an overall trustworthiness weight for the observation. These weights are then used in a standard Dempster-Shafer framework \cite{shafer1976mathematical} to fuse the observations reported by different users, and to compute the trustworthiness for the values of each environmental factor. Table~\ref{tab:1} summarizes the important notations used throughout the paper.

\begin{table}
\caption{Notations}
\label{tab:1}
\centering
\begin{tabular}{|>{\centering\arraybackslash}m{0.6in}| m{2.3in} |} \hline 
\textbf{Notation} 	& 	\multicolumn{1}{|c|}{\textbf{Description}} \\ \hline \hline
$T_{e,t}(u_n)$		& 	The entity trust of user $u_n$ during epoch $t$ \\ \hline 
$T^k_{d,t}(a_l)$ 	& 	The data trustworthiness for environmental factor $f_k$ in area $a_l$ during epoch $t$ \\ \hline 
${OB}^{l,k,i}_{n,t}$ 	& 	The $i$-th observation of user $u_n$ about factor $f_k$ in area $a_l$ during epoch $t$ \\ \hline 
$b_{n,t}$ 			& 	The contributing behavior of user $u_n$  \\ \hline 
$M^{l,k}_{n,t}$ 		& 	The total number of observations that user $u_n$ makes on factor $f_k$ in area $a_l$ during epoch $t$ \\ \hline 
$E^{l,k}_{n,t}$ 		& 	The set of binary evaluations that HAMS makes about observations of user $u_n$ on factor $f_k$ in area $a_l$ during epoch $t$ \\ \hline 
$e^{l,k,i}_{n,t}$ 		& 	The evaluation that HAMS makes about $i$-th observation of user $u_n$ on factor $f_k$ in area $a_l$ during epoch $t$ \\ \hline 
$f_N$ 			& 	The probability of false negatives in HAMS's evaluations \\ \hline 
$f_P$ 			& 	The probability of false positives in HAMS's evaluations \\ \hline 
$S_n$ 			& 	The score of user $u_n$  \\ \hline 
$t_{update}$ 		& 	The time length of trust update interval \\ \hline 
$C_U$ 			& 	The set of user classes \\ \hline 
$C_F$ 			& 	The set of environmental factor classes \\ \hline 
${\mathcal{N}}_t\mathrm{(}\mathbb{C}\mathrm{)}$ 	& 	The number of HAMS's evaluations on user $u_n$ sharing correct observations during epoch $t$ \\ \hline 
${\mathcal{N}}_t(\mathbb{W})$ 		& 	The number of HAMS's evaluations on user $u_n$ sharing wrong observations during epoch $t$ \\ \hline 
${\mu }_L\left(.\right)$ 	& 	The spatial context weight function  \\ \hline 
${\mu }_T\left(.\right)$ 	& 	The temporal context weight function \\ \hline 
${\mu }_C\left(.\right)$ 	& 	The semantic context weight function \\ \hline 
$w\left({OB}^{l,k,i}_{n,t}\right)$ 	& 	The overall trustworthiness weight assigned to the $i$-th observation of user $u_n$ about factor $f_k$ in area $a_l$ during epoch $t$. \\ \hline 
\end{tabular}
\end{table}

\subsection{Entity Trust}
\label{sec:3.1}
In the system model described in Sect.~\ref{sec:2}, the entities are the users who provide observations about the environmental factors. HAMS has to estimate the entity trust $T_{e,t}(u_n)$ by judging the users' contributing behavior over time. In order to perform this estimation, we propose that each user $u_n$ be given a probabilistic score $S_n$ which indicates $u_n$'s inclination towards sharing correct/falsified observations. As the history of observations gathered from $u_n$ grows larger, HAMS incrementally updates the probability distribution of $S_n$. At the end of each epoch $t$, the entity trust $T_{e,t}(u_n)$ can then be updated as the mean of $u_n$'s score, which is computed based on the latest statistical distribution derived from the entire history of interactions. We discuss the details of this entity trust computation process in two parts: \textit{learning users' scores} and \textit{updating the entity trusts}.

\subsubsection{User Score Computation via Bayesian Learning}
\label{sec:3.1.1}

In this section, we discuss how a user's score $S_n$ can be learned by the system from its rounds of interactions with user $u_n$. We define $S_n$ as the probability of user $u_n$ sharing correct observations, i.e.,

\begin{equation}
\label{eq:3}
S_n\mathrm{=}\mathbb{P}\left[b_n=\mathbb{C}\right].
\end{equation}

\par It is further assumed that\textbf{ }$S_n$ takes values from a discrete set of $J$ levels $\mathbb{S}\mathrm{\stackrel{def}{=}}\mathrm{\{}s_1,\dots ,s_J\}$. HAMS can incrementally estimate $S_n$ based on the binary evaluations of observations received from user $u_n$ by the end of each epoch $t$. In particular, the evaluation set is empty initially, and accordingly we assume that the initial value of ${\mathbb{P}}^0\left[S_n=s_j\right]$ is equal to $\frac{1}{J}$, i.e., it follows a uniform distribution (since in the beginning, no observations are available disclosing the user's behavior). As time progresses, a history begins to build up for each user, and we may use standard Bayesian updating rule recursively to update the conditional distribution of $S_n$: 

\begin{equation}
\label{eq:4}
\begin{split}
{\mathbb{P}}^t\left[S_n=s_j|E_{n,t}\right]=  \frac{{\mathbb{P}}^{t-1}\left[S_n=s_j\right]\times {\mathbb{P}}^t\left[E_{n,t}|S_n=s_j\right]}{\sum_{s_j\in \mathbb{S}}{{\mathbb{P}}^{t-1}\left[S_n=s_j\right]\times {\mathbb{P}}^t\left[E_{n,t}|S_n=r_j\right]}} ,
\end{split}
\end{equation}

\noindent where ${\mathbb{P}}^t\left[E_{n,t}|S_n=s_j\right]$ is calculated from Eq.~\ref{eq:5} in Theorem~\ref{th:1}.

\begin{theorem}
\label{th:1}
Define $M\mathrm{\stackrel{def}{=}}\sum_{l,k}{M^{l,k}}$. Also let ${\mathcal{N}}_t\mathrm{(}\mathbb{C}\mathrm{)}$ (resp., \linebreak ${\mathcal{N}}_t(\mathbb{W})$) represent the number of HAMS's evaluations on user $u_n$ sharing correct (resp., false) observations during epoch  $t$\textit, where ${\mathcal{N}}_t\left(\mathbb{C}\right)+{\mathcal{N}}_t\left(\mathbb{W}\right)=M$. It then holds that:
\end{theorem}

\begin{equation}
\label{eq:5}
\begin{split}
{\mathbb{P}}^t\left[E_{n,t}|S_n=s_j\right]=s_j\times {\left(f_P\right)}^{{\mathcal{N}}_t(\mathbb{W})}\times {\left({1-f}_P\right)}^{{\mathcal{N}}_t\mathrm{(}\mathbb{C}\mathrm{)}}+ \\
\left(1-s_j\right)\times {\left(f_N\right)}^{{\mathcal{N}}_t\mathrm{(}\mathbb{C}\mathrm{)}}\times {\left({1-f}_N\right)}^{{\mathcal{N}}_t(\mathbb{W})}.
\end{split}
\end{equation}

\renewenvironment{proof}{{\bf{Proof.} }}{\hfill $\square$ \\} 
\noindent \begin{proof}
See Appendix A.
\end{proof}

\subsubsection{Updating Entity Trust}
\label{sec:3.1.2}
 At the end of each epoch $t$, the value of entity trust $T_{e,t}(u_n)$ for user $u_n$ is updated as the mean value of $S_n$. Using the latest conditional distribution of the user's score, $T_{e,t}(u_n)$ can be calculated as follows:
 
 \begin{equation}
 \label{eq:6}
 T_{e,t}\left(u_n\right)=\sum_{s_j\in \mathbb{S}}{s_j\times {\mathbb{P}}^t\left[S_n=s_j|E_{n,t}\right]}.
 \end{equation}

\subsection{Data Trustworthiness}
\label{sec:3.2}

In the proposed scenario, data trustworthiness $T^k_{d,t}(a_l)$ indicates how trustworthy HAMS's claim about the healthiness of area $a_l$ according to factor $f_k$ is. We determine $T^k_{d,t}(a_l)$ by taking into account the following four critical elements: the trust in data contributor (i.e., entity trust), the semantic trustworthiness of the environmental factor and its observer, the consistency between the observer's home location and the observed area, and finally, the temporal freshness of the generated observation. Our proposed method for the computation of data trustworthiness consists of two phases: \textit{data observation weight assignments} and \textit{observation aggregations}.

\subsubsection{Data Observation Weight Assignment}
\label{sec:3.2.1}

Data trustworthiness is inferred from aggregating all the observations contributed by different users. Some of these observations are correct, and some of them are wrong either due to users' malice or simply because of inaccurate sensor readings. The purpose of the weight assignment phase is to weigh the contributions received from different users considering both: the context in which each observation has been created as well as the entity trust of the observer $T_{e,t}\left(u_n\right)$ (as evaluated by the proposed entity-centric trust model). In what follows, we introduce the contextual elements underlying data trustworthiness, and discuss how to factor these elements into weighing the observations. We consider three contextual elements associated with each submitted observation: \textit{spatial}, \textit{temporal}, and \textit{semantic context}:

\textbf{ Spatial Context:} It can be argued that a sensing middleware is more likely to collect the most informative data when the users are in their home areas, i.e., the areas the user visits more frequently, rather than other areas. This can be mainly attributed to the fact that observations stemming from a user's home area are sufficiently localized so that the sensors can be calibrated towards accuracy. In fact, many automated sensor calibration techniques leverage on machine learning, and the number of contributed measures by the sensor needs to be high enough to overcome the low accuracy of the ordinary sensors. Armed with this understanding, Eq.~\ref{eq:7} formulates a simple weighing function to capture the impact of the spatial context on the observations made by user $u_n$:

\begin{equation}
\label{eq:7}
{\mu }_{L}\left({OB}^{l,k,i}_{n,t}\right)=\left\{ \begin{array}{c}
{\mathbb{H}}_{SP},\ \ a_l=a_{home(u_n)} \\ 
{\mathbb{L}}_{SP},\ \ a_l\neq a_{home(u_n)} \end{array},
\right.
\end{equation}

\noindent where ${\mathbb{H}}_{SP}$ and ${\mathbb{L}}_{SP}$ are system-specific parameters indicating high (resp., low) weights assigned to observations coming from $u_n$'s home (resp., non-home) areas.

\textbf{Temporal Context:} The state of an event in a system may change over time, and an event may only be relevant for a limited portion of time. Hence, the observations that are contributed more closely to the end of each $t_{update}$ interval are more likely to be accurate. Let $\tau ({OB}^{l,k,i}_{n,t})$ denote the timestamp associated with a given observation. The weighing function ${\mu }_T:\ {\mathbb{R}}^+\to [0,1]$ as given by Eq.~\ref{eq:8} uses simple parametric thresholds to give more weights to more temporally fresh observations: 

\begin{equation}
\label{eq:8}
{\mu }_T\left(\tau ({OB}^{l,k,i}_{n,t})\right)=
\begin{cases}
{\mathbb{H}}_T,\ \ t_{update}-\tau ({OB}^{l,k,i}_{n,t})\mathrm{\ }<t_1 \\ 
{\mathbb{M}}_T,\ \ t_1<t_{update}-\tau ({OB}^{l,k,i}_{n,t})\mathrm{\ }<t_2 , \\ 
{\mathbb{L}}_T,\ \ {t_2<t}_{update}-\tau ({OB}^{l,k,i}_{n,t})\mathrm{\ }
\end{cases}
\end{equation}

\noindent where $t_1$ and $t_2$ are defined as two interval indicators that can be set according to the characteristics of each factor. Also, ${\mathbb{H}}_T$, ${\mathbb{M}}_T$, and ${\mathbb{L}}_T$ are system-specific parameters denoting high, medium, and low grades to be given to observations based on their recency.

\begin{table*}[!tbh]
\caption{Semantic Context}
\label{tab:2}
\centering
{\small
\hfill{}
\begin{tabular}{|>{\centering\arraybackslash}m{2.3in}| >{\centering\arraybackslash}m{1.9in} |>{\centering\arraybackslash}m{1.9in}|}
\hline
\backslashbox[6.25cm] {\textbf{User Class}}{\textbf{Factor Class}} 	&		 \textbf{Environmental Health (e.g.,air pollution, temperature,humidity, noise pollution, etc.)} 	& 	\textbf{Urban Accessibility (e.g., facility maturity, infrastructure, sidewalk condition, etc.)} \\ \hline \hline
Healthcare professionals (e.g., physicians, nurses, EMTs, etc.) 		& 		${\mathbb{H}}_{SM}$ 	& ${\mathbb{M}}_{SM}$ \\ \hline 
Vulnerable groups (elderly, disabled, patients, carers, etc.) 		& 		${\mathbb{M}}_{SM}$ 	& ${\mathbb{H}}_{SM}$ \\ \hline 
Government professionals (e.g., CHOs) 						& 		${\mathbb{H}}_{SM}$ 	& ${\mathbb{M}}_{SM}$ \\ \hline 
Regular people 										& 		${\mathbb{L}}_{SM}$ 	& ${\mathbb{L}}_{SM}$ \\ \hline 
\end{tabular}}
\hfill{}
\end{table*}

\textbf{Semantic Context:} In order to capture the semantic trustworthiness of each user class in relation with each class of environmental factors, we introduce a tabular function ${\mu }_C:C_U\times C_F\to [0,1]$. For a given pair of user-factor class ($c_p,c_q$), the function ${\mu }_C\mathrm{(}c_p,c_q\mathrm{)}$ indicates the inherent suitability of a user belonging to class $c_p$ for generating observations on a factor of class $c_q$. This table is meant to be preset based on expert opinions and using empirical experiments. For example, Table~\ref{tab:2} is a very rough example of how the function ${\mu }_C(.)$ can be formulated for two classes of environmental factors concerning environmental health and urban accessibility. The high, medium, and low weight values for the semantic correlations are shown parametrically by ${\mathbb{H}}_{SM}$, ${\mathbb{M}}_{SM}$, and ${\mathbb{L}}_{SM}$, respectively. For instance, when it comes to contributing observations about urban health, government professionals that monitor the environment receive the highest weight followed by emergency management and healthcare professionals or responders that generally care for the public health. On the other hand, patients, vulnerable groups of users, and their carers who are more conscious towards welfare-related issues, are supposedly more eligible to provide information on the accessibility factors in each area. It should be noted however that semantic context only provides a startup template. In fact, although other observers may initially have less trust associated with the data they provide, but as discussed in Sect.~\ref{sec:3.1}, this initial bias may be dominated by the user scores, given that these scores can be adapted gradually based on the content the users create.

\par Overall, once the context-related weights are assigned to an observation, we use Eq.~\ref{eq:9} as a simple convex combination (with $\alpha $,$\ \beta $, and $\theta $ as coefficients) to integrate the impact of contextual properties of a received observation with the entity trust of the observer. In particular, the overall trustworthiness weight $w\left({OB}^{l,k,i}_{n,t}\right)$ is assigned to the $i$-th observation of user $u_n$ about factor $f_k$ in area $a_l$ during epoch $t$:

\begin{equation}
\label{eq:9}
\begin{split}
w\left({OB}^{l,k,i}_{n,t}\right)=\alpha \left(T_{e,t}\left(u_n\right).{\mu }_C\left({OB}^{l,k,i}_{n,t}\right)\right)+\beta \ {\mu }_L\left({OB}^{l,k,i}_{n,t}\right) \\
+\theta \ {\mu }_T\left({OB}^{l,k,i}_{n,t}\right), \alpha +\beta +\theta =1.
\end{split}
\end{equation}

\subsubsection{Observation Aggregation}
\label{sec:3.2.2}
In this step, the weights of all observations about area $a_l$ are aggregated to assess $T^k_{d,t}(a_l)$. This value gets updated periodically after each $t_{update}$. Considering the fact that the observations about area $a_l$ can be contradictory due to malicious behavior of the users, it is essential to utilize a method that can handle the aggregation of contradicting uncertain evidence. We leverage on Dempster-Shafer Theory of evidence (DST) to aggregate different observations from different users about the environmental factors in each area. DST is a general framework for reasoning with uncertainty. The theory allows one to combine evidence from different sources and arrive at a degree of belief (represented by a mathematical object called \textit{belief function}) that takes into account all the available evidence.

\par DST's formalism starts with a set of possibilities under consideration, for instance the numerical values of a variable. In DST's jargon, this set is called \textit{frame of discernment}, and is similar in concept to the notion of state space in probability. In our case, for each factor $f_k$, we need to define a frame of discernment which we denote by ${\mathrm{\Omega }}_k$. In particular, we assume that the entire range of possible numerical values for each observation can be quantized into a total of ${\mathrm{\Gamma }}_k$ real intervals. HAMS is then able to map the value of each reported observation ${OB}^{l,k,i}_{n,t}$ to its corresponding interval from within ${\mathrm{\Omega }}_k$, where

\begin{equation}
\label{eq:10}
\begin{split}
\mathrm{\Omega }_k\mathrm{\stackrel{def}{=}}\biggl\{{\omega }_{k,1}\mathrm{\stackrel{def}{=}} [-\infty ,{\mathrm{r}}_{k,1}),{\omega }_{k,2}\mathrm{\stackrel{def}{=}}[{\mathrm{r}}_{k,1}{\mathrm{r}}_{k,2}),\dots,  \\ 
{\omega }_{k,{\mathrm{\Gamma }}_k}\mathrm{\stackrel{def}{=}}[{\mathrm{r}}_{k,{\mathrm{\Gamma }}_k}+\infty)\biggr\}.
\end{split}
\end{equation}

\par This quantization can be done using expert opinion and in a meaningful fashion for each factor; for instance, the range of observations about air pollution factors can be quantized to levels indicating different health grades such as: ``healthy'', ``borderline'', ``unhealthy for sensitive groups'', ``unhealthy'', ``very unhealthy'' and ``hazardous''.

\par Now, a \textit{hypothesis} in DST is represented by a subset of the frame of discernment. For each ${\mathrm{\Omega }}_k$, there are $2^{{\mathrm{\Gamma }}_k}$ hypotheses (including the hypotheses null $\emptyset $ and universal ${\mathrm{\Omega }}_k$). We denote this set of hypotheses by power set $2^{{\mathrm{\Omega }}_k}$. In a first step, subjective probabilities (\textit{masses}) are assigned to all subsets of the frame; usually, only a restricted number of hypotheses (namely, \textit{focal elements}) have non-zero mass. In our case, if the received observation ${OB}^{l,k,i}_{n,t}$ from user $u_n$ happens to belong to interval ${\omega }_{k,\gamma }\in {\mathrm{\Omega }}_k$ ($\gamma =1,2,\dots ,{\mathrm{\Gamma }}_k$), we exploit the observation weights from Eq.~\ref{eq:11} to assign mass to every hypothesis $h\in 2^{{\mathrm{\Omega }}_k}\ $as follows: 

\begin{equation}
\label{eq:11}
m^{l,k,i}_{n,t}\left(h\right)=
\begin{cases}
w\left({OB}^{l,k,i}_{n,t}\right), & \ {h=\omega }_{k,\gamma } \\ 
1-w\left({OB}^{l,k,i}_{n,t}\right), & \ h{\mathrm{=}\mathrm{\Omega }}_k\  \\ 
0, & \ Otherwise
\end{cases}.
\end{equation}

\par In fact, since each user can only contribute a single value for a given factor $f_k$, and there is no overlap between the quantization intervals, the only non-universe hypothesis that can have positive mass is the interval entailing the numerical value of ${OB}^{l,k,i}_{n,t}$. Also, in DST, the mass assigned to the universal set ${\mathrm{\Omega }}_k$ refers to the proportion of evidence that can't be assigned to any of the other hypotheses.

\par Now, at the end of each update epoch $t$ when different users have expressed their beliefs over the frame, \textit{Dempster's rule of combination} can be used by HAMS to compute the combination (called the \textit{joint mass}) of all sets of masses corresponding to any interval ${\omega }_{k,\gamma }\in {\mathrm{\Omega }}_k$ of numerical values for each factor $f_k$ in area $a_l$. For simplicity, in order to explain the DST's fusion operator, we consider the combination of two sets of masses $m^{l,k,i}_{n,t}$ and $m^{l,k,j}_{\acute{n},t}$ corresponding respectively to the masses assigned to the $i$-th observation of user $u_n$, and the $j$-th observation of user $u_{\acute{n}}$. According to DST's rule, for each interval of numerical values ${\omega }_{k,\gamma }\in {\mathrm{\Omega }}_k,\gamma \in \left\{1,2,\dots ,{\mathrm{\Gamma }}_k\right\},$ we have that:

\begin{equation}
\label{eq:12}
\begin{split}
m^{l,k,i}_{n,t}\left({\omega }_{k,\gamma }\right)\bigoplus m^{l,k,j}_{\acute{n},t}\left({\omega }_{k,\gamma }\right)= \hspace{105pt} \\
\hspace {5pt} \frac{\sum_{h\ \cap \acute{h}=\ {\omega }_{k,\gamma }}{m^{l,k,i}_{n,t}\left(h\right)\times m^{l,k,j}_{\acute{n},t}\left(\acute{h}\right)}}{1-\ \sum_{\ h\cap \acute{h}=\ \emptyset }{m^{l,k,i}_{n,t}\left(h\right)\times m^{l,k,j}_{\acute{n},t}\left(\acute{h}\right)}}, \ h,\acute{h}\in 2^{{\mathrm{\Omega }}_k}.
\end{split}
\end{equation}

\par In fact, DST derives a common shared belief between multiple users, while ignoring all the conflicting (non-shared) belief through a normalization factor. An advantage of DST is that its fusion operator can be interpreted as an approximate generalization of Bayes' rule, in which (unlike traditional Bayesian methods) the priors and conditionals need not be pre-determined.

\par Although we have expressed DST's fusion operator in a pairwise fashion, but the operator $\bigoplus $ can be applied repeatedly to fuse any number of observations. In fact, starting from the outset, it suffices to combine the mass of a third observation with the joint mass computed for the first two. The final joint mass obtained in this fashion indicates the trustworthiness of HAMS's claim that the value of $f_k$ for area $a_l$ falls within the range ${\omega }_{k,\gamma }\in {\mathrm{\Omega }}_k$. More formally, since each user $u_n$ contributes a total of $M^{l,k}$ observations over the course of an epoch $t$, we use the symbol $m^{l,k}_{n,t}$ to denote the joint mass derived from fusing all observations reported by $u_n$. Now, the combined mass for user $u_n$ is computed as follows: 

\begin{equation}
\label{eq:13}
\begin{split}
m^{l,k}_{n,t}\left({\omega }_{k,\gamma }\right)\mathrm{\stackrel{def}{=}}{\bigoplus }^{M^{l,k}}_{i=1}m^{l,k,i}_{n,t}\left({\omega }_{k,\gamma }\right),{\omega }_{k,\gamma }\in {\mathrm{\Omega }}_k, \\
\forall \ \gamma \in \left\{1,2,\dots ,{\mathrm{\Gamma }}_k\right\}.
\end{split}
\end{equation}

\par Accordingly, we use the symbol $m^{l,k}_t$ to denote the joint mass derived from fusing the observations from the users who contributed observations about the factor $f_k$ in area $a_l$ during epoch $t$. $m^{l,k}_t$ can be computed from ${\left\{m^{l,k}_{n,t}\right\}}_{n=1,\dots ,N}$ as follows: 

\begin{equation}
\label{eq:14}
\begin{split}
m^{l,k}_t\left({\omega }_{k,\gamma }\right)\mathrm{\stackrel{def}{=}}{\bigoplus }^{}_{n \in U\left(t\right)}m^{l,k}_{n,t}\left({\omega }_{k,\gamma }\right), {\ \ \ \omega }_{k,\gamma }\in {\mathrm{\Omega }}_k, \\
\forall \ \gamma \in \left\{1,2,\dots ,{\mathrm{\Gamma }}_k\right\},
\end{split}
\end{equation}

\par where $U\left({t}\right) \subseteq U$ denotes the subset of users who actually contribute observations during the $t$-th epoch.

\par Finally, the data trustworthiness $T^k_{d,t}\left(a_l\right)$ associated with factor $f_k$ is a vector containing the joint masses associated with all the intervals  ${\omega }_{k,\gamma }\in {\mathrm{\Omega }}_k$; i.e., 

\begin{equation}
\label{eq:15}
\begin{split}
T^k_{d,t}\left(a_l\right)={\left\{m^{l,k}_t\left({\omega }_{k,\gamma }\right)\right\}}_{\ \gamma =1,2,\dots ,{\mathrm{\Gamma }}_k}.
\end{split}
\end{equation}


\section{Performance Evaluation}
\label{sec:4}
In this section, performance measurements using an in-house simulator are provided using different scenarios to demonstrate the system’s properties. After introducing the simulation setting in Sect.~\ref{sec:4.1}, we present the results in two sections: Sect.~\ref{sec:4.2} demonstrates the convergence and the accuracy properties of our scheme as well as its resilience against changes in users’ behavior. Sect.~\ref{sec:4.3} is devoted to a comparative evaluation of our proposed method against \cite{al2017trust}, which is a recent trust management scheme, proposed specifically for health IoT applications.

\subsection{Parameter Settings}
\label{sec:4.1}
For the sake of experiments, we consider an IoT environment comprised of $N=40$ users distributed uniformly in $L=100$ areas. Given that our scheme prescribes an identical procedure for all environmental factors, for simplicity, we assume that only one environmental health factor, e.g., the concentration of PM2.5 in the air, is being observed by the users. Each user is able to query HAMS about the health factor of one area before entering the area, and contributes an observation about this health factor. Again, for simplicity, we consider only two levels of quantization for the values of the factor under observation $({\mathrm{\Gamma }}_k=2)$, indicating very coarsely whether it is currently within a healthy range or not.

\par The percentage of malicious users (PMU) in our system ranges from 0 to 70\% to demonstrate the effect of malicious behavior of the users on the performance of our trust management system. The simulations are conducted using three trust score levels, i.e., $J=3$. As such, the score set consists of three elements $\{s_1,s_2,s_3\}$, which are proportionally set to indicate the scores of the malicious, ignorant, and good users. The trust scores of all users are initially set to 0.5 (i.e., ignorance), and they should eventually converge to their true score levels.

\par The coefficients used in Eq.~\ref{eq:9} for assigning weights to each observation are set as $\alpha =0.7$ and $\beta =0.3$. As for $\theta $ (the temporal context coefficient), we assume for simplicity that the environment has a long timescale model so that an unhealthy area stays unhealthy throughout the simulation; therefore, the temporal context is not used for assessing data trustworthiness, and we set $\theta =0$. The values chosen for the parametric weights associated with the semantic correlation function ${\mu }_C(.)$ and the spatial context ${\mu }_L(.)$ (used in Eq.~\ref{eq:7} for data trustworthiness calculation) are demonstrated in Table~\ref{tab:3}. The false positive and false negative parameters in our system are set as $f_N=f_P=0.2$. A more detailed discussion on the impact of these two parameters is presented in the next subsection.

\par The TM module in HAMS calculates both entity trusts and data trustworthiness after each $t_{update}$ elapses; accordingly, in our plots, the time axis shows the progression of the algorithm in terms of  $t_{update}$-length epochs.

\begin{table}
\centering
\caption{Parameters for Performance Evaluation}
\label{tab:3} 
\begin{tabular}{|m{0.5in} m{0.5in} m{0.8in} m{0.7in}|} \hline
\textbf{Parameter} 		& 	\textbf{Value} 	& 	\textbf{Parameter} 					& 	\textbf{Value} \\ \hline
$N$ 					& 	$40$ 		& 	$\alpha $ 							& 	$0.7$ \\ 
$L$ 					& 	$100$ 		& 	$\beta $ 							& 	$0.3$ \\  
${\mathrm{\Gamma }}_k$ & 	$2$ 			& 	$\{s_1,s_2,s_3\}$ 					& $\{0.05,0.5,0.95\}$ \\ 
$J$ 					& 	$3$ 			& 	$\{{\mathbb{H}}_{SP},{\mathbb{L}}_{SP}\}$ 	& $\left\{1,0.4\right\}$ \\ 
$f_N$, $f_P$  &  0.2 & $\{{\mathbb{H}}_{SM},{\mathbb{M}}_{SM},{\mathbb{L}}_{SM}\}$ & $\left\{1,0.8,0.6\right\}$ \\  \hline
\end{tabular}
\end{table}

\subsection{Convergence, Accuracy and Resiliency}
\label{sec:4.2}
In this section, we evaluate the convergence behavior and the accuracy of our proposed TM scheme as well as its resiliency by exploring the impact of changes in users’ behavior. We also evaluate the accuracy of the assessed trust against the ground truth as the population of the malicious users increases.

\par
\begin{figure}
\centering
\includegraphics*[width=0.5\textwidth]{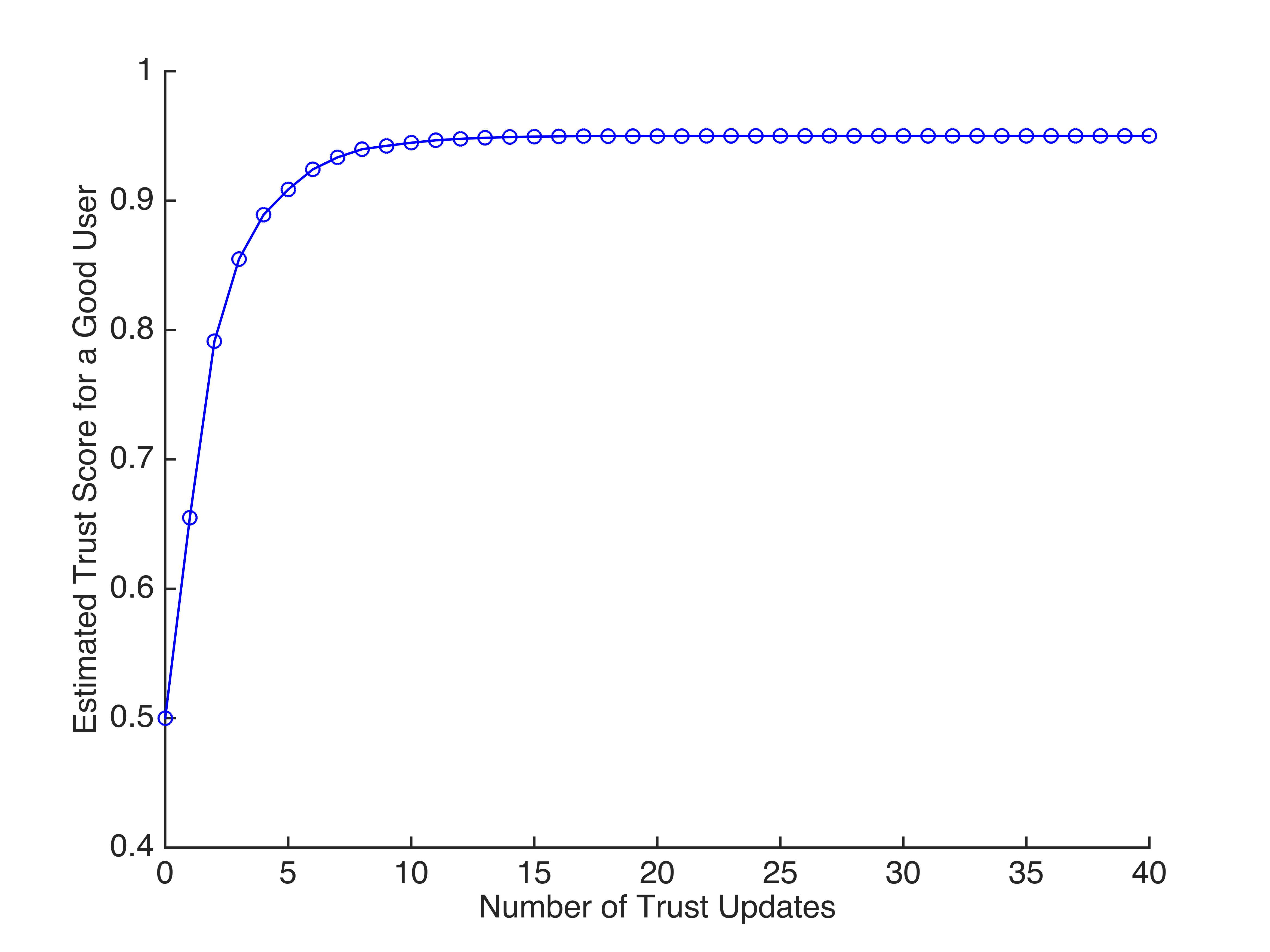}
\caption{Convergence of the Estimated Trust Score for a “Good” User}
\label{fig:3} 
\end{figure}

\begin{figure}
\centering
\includegraphics*[width=0.5\textwidth]{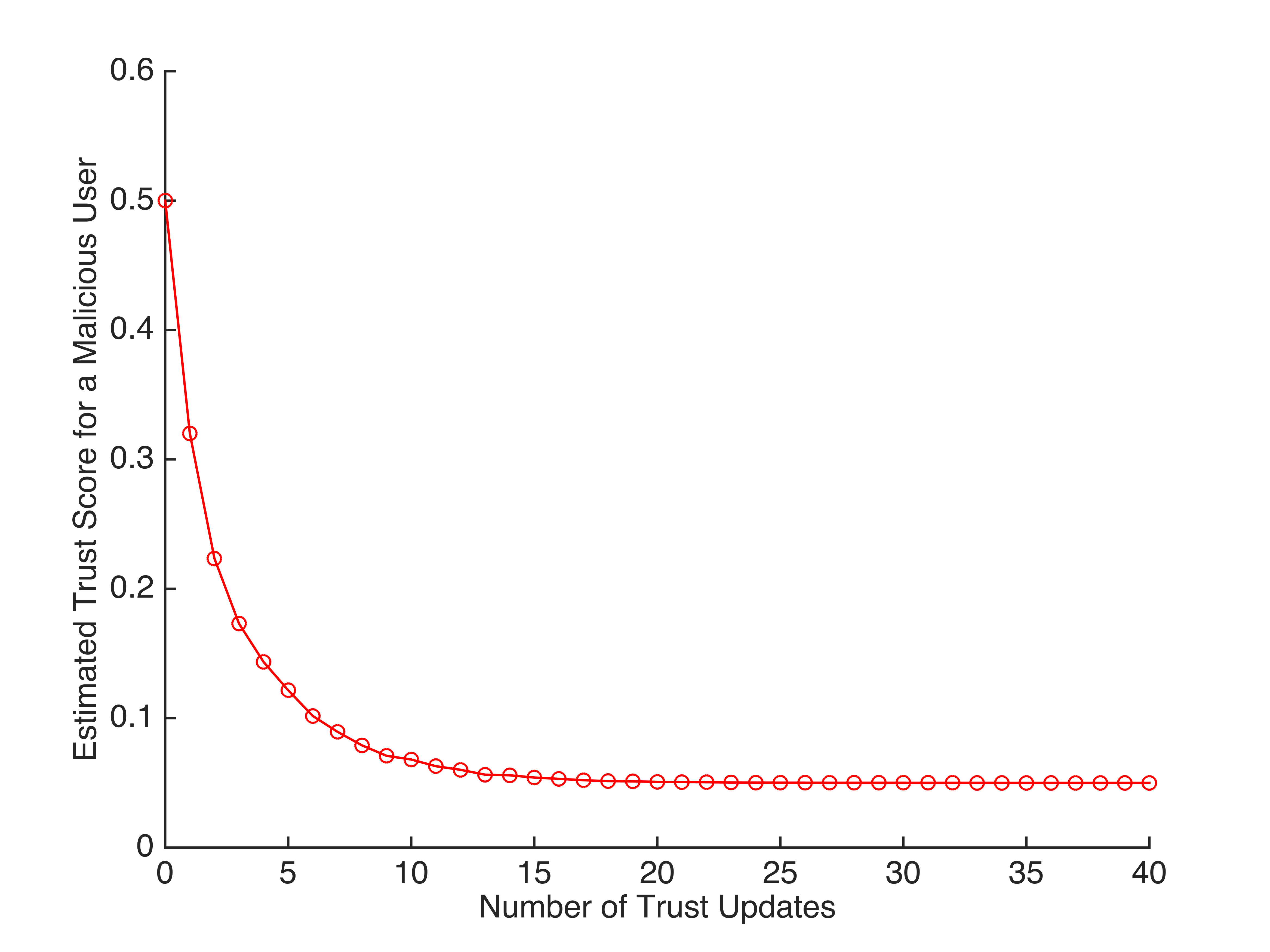}
\caption{Convergence of the Estimated Trust Score for a “Malicious” User}
\label{fig:4} 
\end{figure}

 Fig.~\ref{fig:3} illustrates the evolution of the estimated entity trust score for a “good” user randomly selected from the population. For this experiment, the parameters $f_N$ and $f_P$ are set according to Table~\ref{tab:3}. We see that as time progresses and more observations are gathered, the entity trust value of a “good” user converges around the expected score for the good users which is $0.95$. Similarly, Fig.~\ref{fig:4} shows how the estimation of entity trust score for a malicious user progresses with time. Again, we set the simulation parameters according to Table~\ref{tab:3}, and the “malicious” user is randomly selected from the population of the users. We see that as time passes, the plot converges around the expected score for the malicious users which is $0.05$.Overall, these plots corroborate the efficacy of the proposed Bayesian learning procedure for estimating the random user trust scores from the history of their contributing behavior. 

\par
\begin{figure}
\centering
\includegraphics*[width=0.5\textwidth]{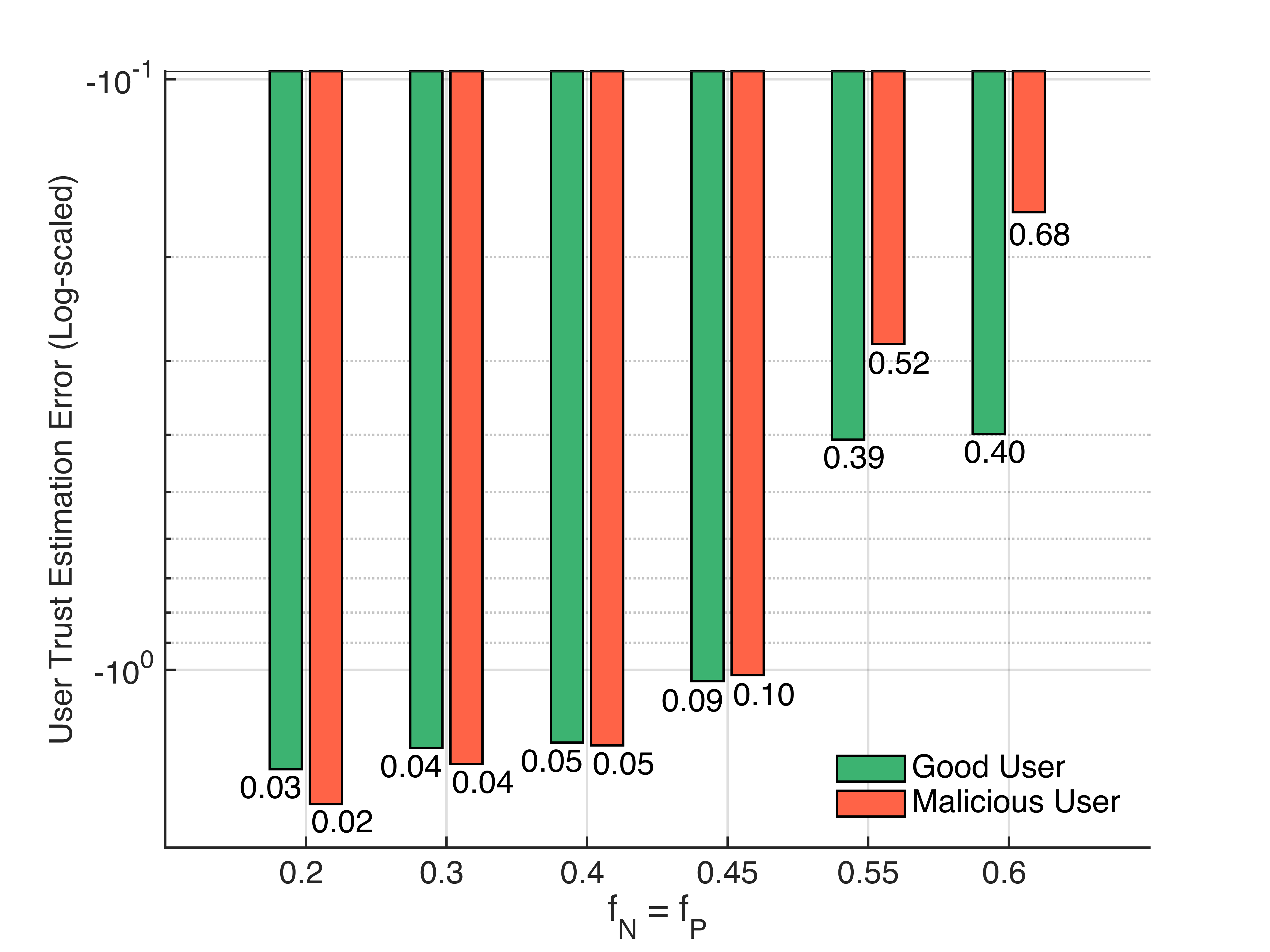}
\caption{User Trust Estimation Error}
\label{fig:5} 
\end{figure}

 Next, We experiment with the impact of different misdetection error probabilities (i.e., false negative/positive parameters denoted by $f_N$ and $f_P$) on our entity trust estimation process. In Fig.~\ref{fig:5}, we compare the differences between the estimated trust score and the expected trust score of randomly selected “malicious” and “good” users, considering different $f_N$ and $f_P$. We name this difference, “user trust estimation error”. We plotted this figure in a log-scale format to demonstrate the results more clearly. As can be seen, the learning process is fairly robust against reasonable values for the misdetection probabilities; however, the performance ultimately suffers when the learning algorithm is equipped with a poor detection capability (e.g., when the values of $f_N$ and $f_P$ are higher than $0.5$, the error value jumps up to $0.48$).  As such, the performance of the entity trust computation method depends largely on the parameters $f_N$ and $f_P$. However, if the system is tuned with reasonably low values of misdetection probabilities, the proposed scheme is very effective in estimating the trust scores of the users, e.g., for the value of $f_N$ and $f_P$ equal to $0.2$, the user trust estimation error is as low as $0.02$.

\par
\begin{figure}
\centering
\includegraphics*[width=0.5\textwidth]{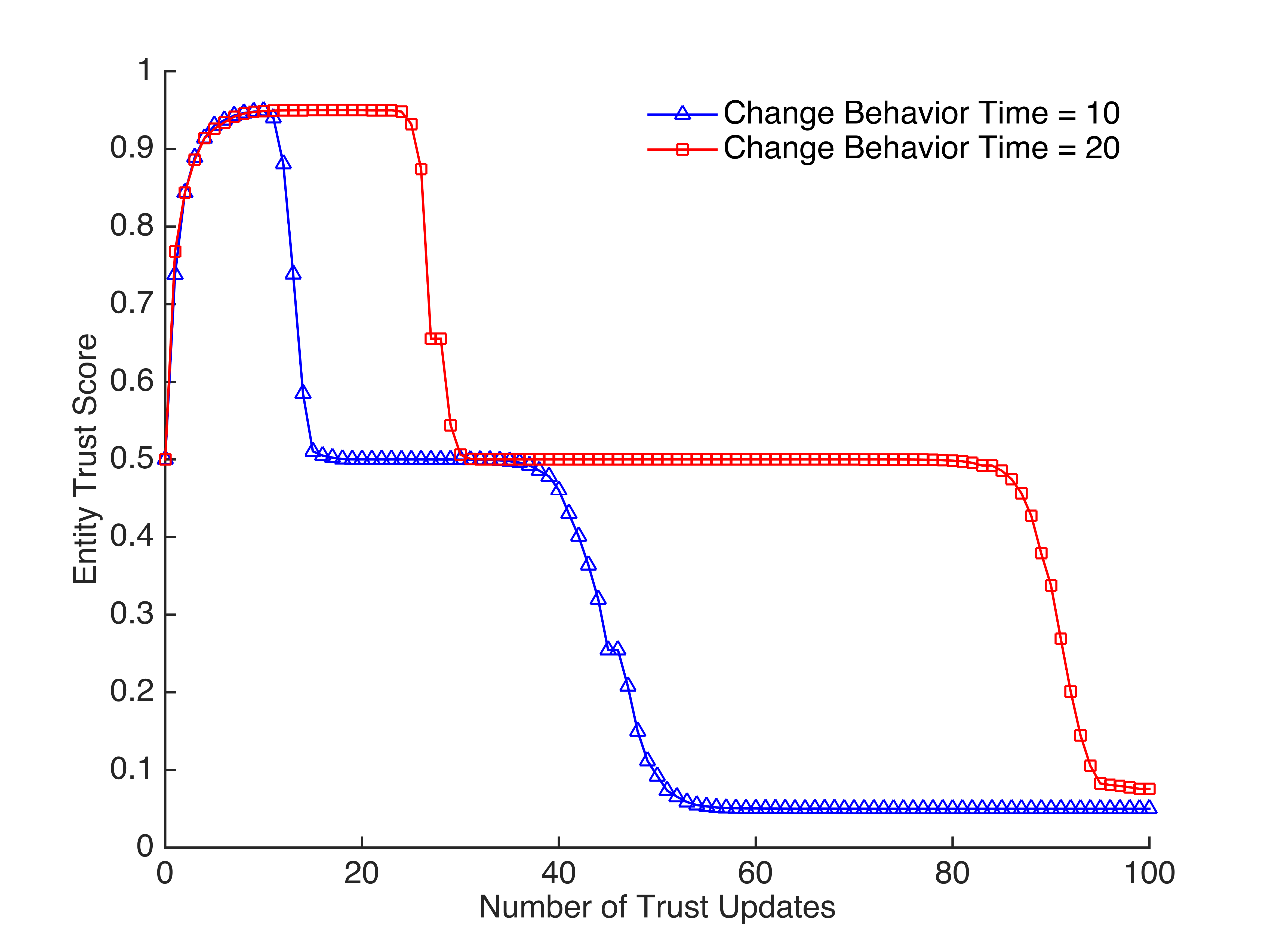}
\caption{Tracking Entity Trust Scores under (User) Behavioral Changes}
\label{fig:6} 
\end{figure}

 To evaluate the resiliency of the proposed TM system, we consider a case in which the user’s behavior changes in the middle of one simulation, e.g., a good user turns into malicious (or vice versa). This experiment is meant to demonstrate the responsiveness of our proposed TM in detecting and tracking such changes. Fig.~\ref{fig:6} illustrates the entity trust of a user whose behavior is subject to sudden change. We consider two scenarios; in the first case, the user’s behavior changes after the $10^{th}$ epoch, and in the second case, the user’s behavior changes after $20$ epochs. In order to demonstrate the impact of this behavioral change, the duration of these experiments is considered to be longer than the setup we mentioned earlier. 

\par In the first scenario, given that the user is not malicious at first, the estimated value for the entity trust score is high in the beginning, i.e., around $0.95$. Once the behavioral change occurs at the $10^{th}$ epoch, the estimated value for the entity trust score drops quickly to the score of ignorant users which is $0.5$. After receiving more observations from this user, the estimation picks up its downward trend till it converges around the score of malicious users (which is $0.05$). Overall, as evidenced by this experiment, the estimation of entity trusts is able to track the change in users’ behavior.

\par In the second scenario, the user’s behavior changes at the $20^{th}$ epoch. Compared to the first case, we can see that it takes more time for the TM module to detect that a behavioral change has occurred. In particular, since the history of previous observations is lengthier in this case, it takes more time for the estimation process to forget the past behavior of the user, and for the recent observations to manifest themselves. In fact, it has taken almost twice as long for the entity trust value to converge around the score of the malicious users (i.e., $0.05$).

\par
\begin{figure}
\centering
\includegraphics*[width=0.5\textwidth]{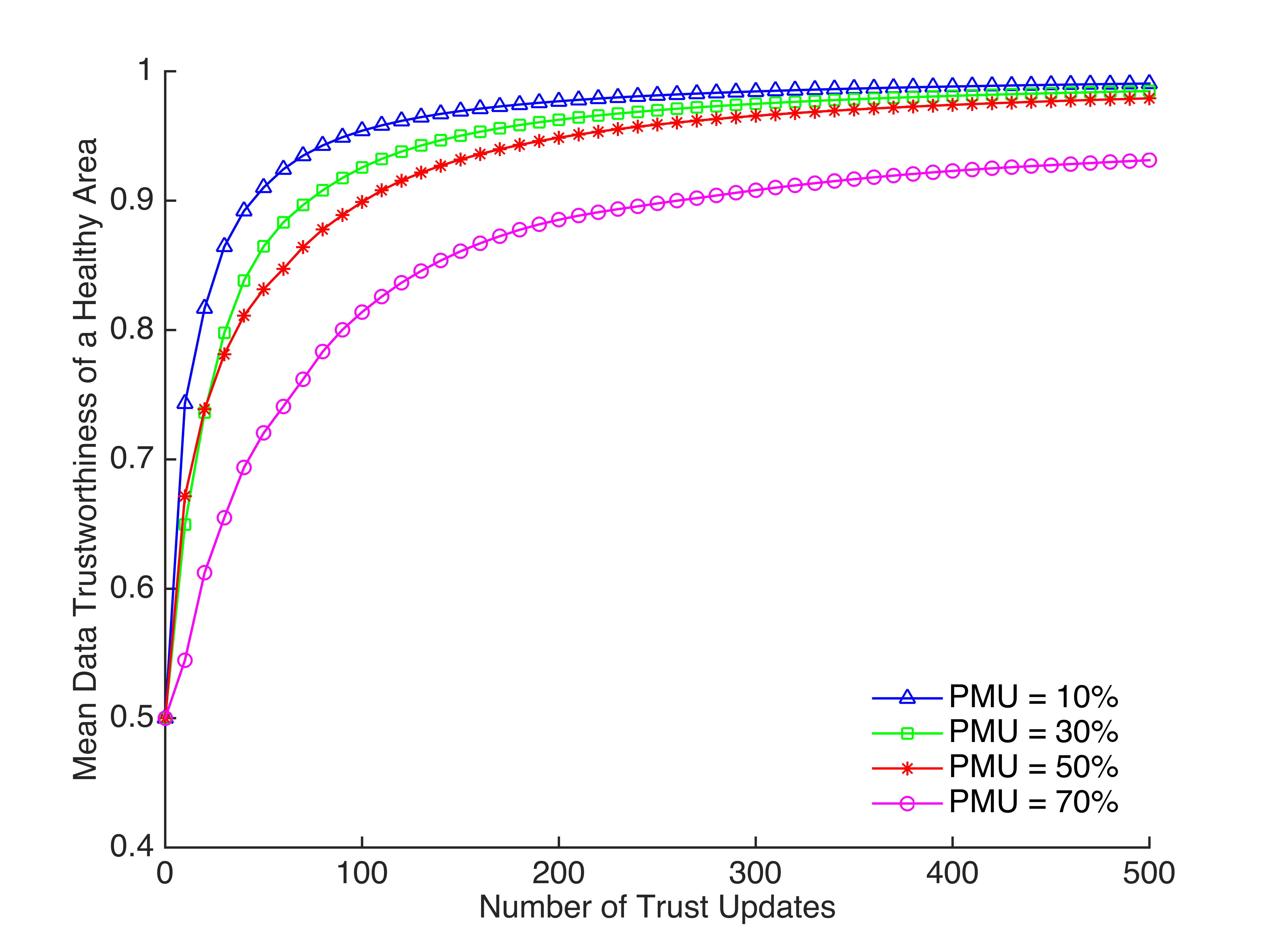}
\caption{The Mean Data Trustworthiness of a Healthy Area}
\label{fig:7} 
\end{figure}

\begin{figure}
\centering
\includegraphics*[width=0.5\textwidth]{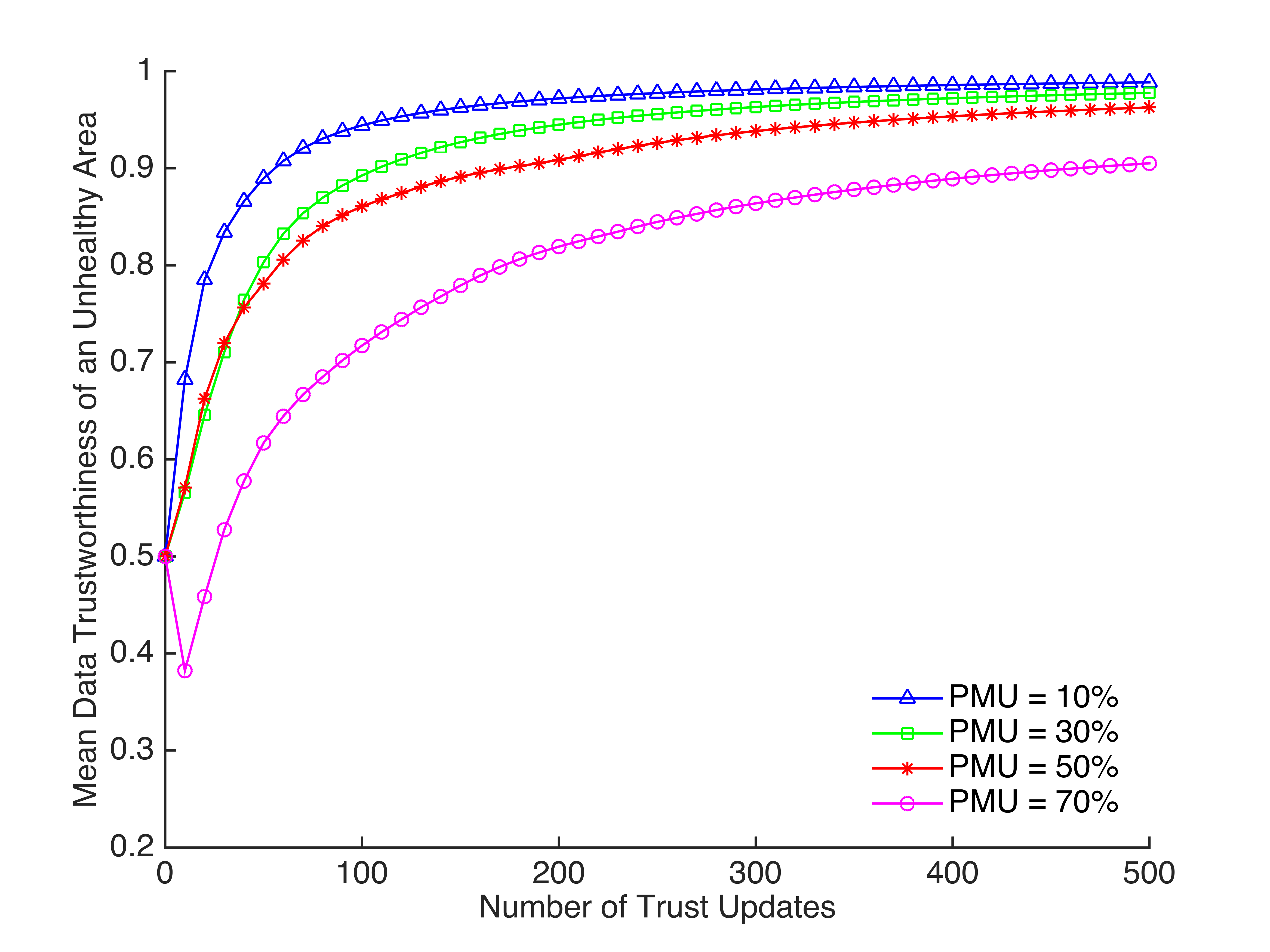}
\caption{The Mean Data Trustworthiness of an Unhealthy Area}
\label{fig:8} 
\end{figure}

 In the next set of experiments, we demonstrate the performance of the proposed scheme for estimating the data trustworthiness. Again, the simulation settings complies with Table~\ref{tab:3}. Fig.~\ref{fig:7}  plots the estimated mean data trustworthiness of HAMS’s claims about the healthiness of a given healthy area vs. time. As can be observed, when only 10\% of the users in our system are malicious, the data trustworthiness is nearly 1, which shows that HAMS could spot the healthy areas almost perfectly. As PMU increases, however, the malicious behavior prevails gradually in the system, generally resulting in a reduction of the data trustworthiness of HAMS’s claims; but still, even under high PMU values, $T^k_{d,t}(a_l)$ converges to $0.8$, (well above $0.5$ as an obvious indecisive threshold), which demonstrates that HAMS’s claims about the healthiness of an area can very well be counted upon as trustworthy. Similarly, Fig.~\ref{fig:8} plots the mean data trustworthiness of HAMS’s claims about the unhealthiness of an unhealthy area vs. time. As can be seen, the value of $T^k_{d,t}(a_l)$ for all PMUs is also well above $0.5$, which indicates that HAMS’s claims about the unhealthiness of a given area are trustworthy. 

\subsection{Comparison with Related Work}
\label{sec:4.3}
In this section, we compare the proposed scheme for the computation of data trustworthiness and entity trust with the recent method presented in \cite{al2017trust}. \cite{al2017trust} uses a weighted sum approach to assess the entity trust of each user. Basically, \cite{al2017trust} compares each user’s observation about an area to the other observations received about this area to estimate the correctness of each observation. For the computation of data trustworthiness, this method assigns a weight to each observation, and finds the average of the observations’ weights for each area. The weight is the entity trust of the contributor; therefore, the data trustworthiness indicates the trustworthiness of HAMS’s claim about the health status of an area inferred from users’ observations.

\par As for entity trust comparisons, we only present the results for the “good” subset of the users (the comparison for “malicious” users is similar). Comparison with respect to data trustworthiness is illustrated for the “healthy” areas considering one factor (the comparison for the “unhealthy” areas is similar). 

\begin{figure}
\centering
\includegraphics*[width=0.5\textwidth]{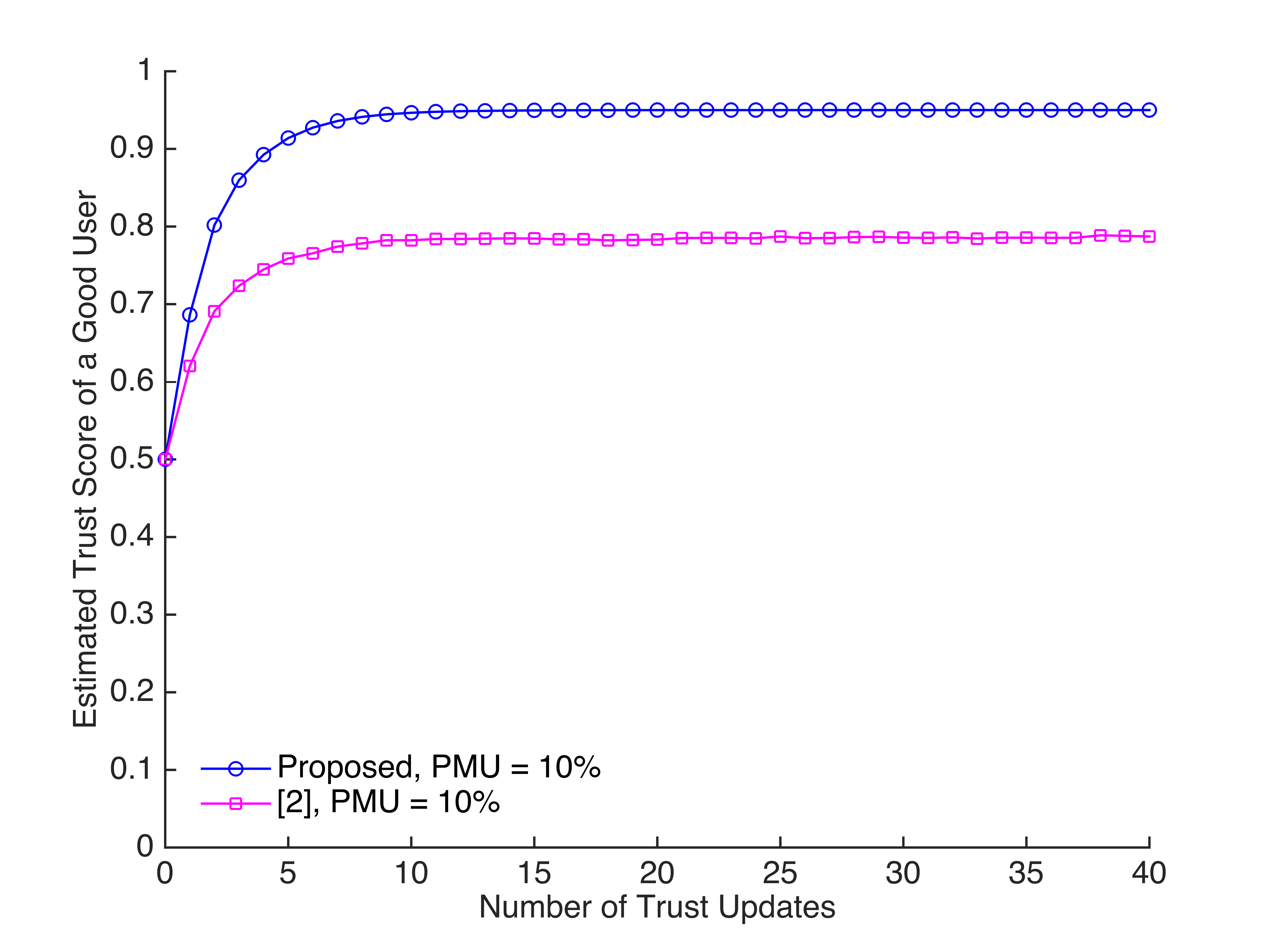}
\caption{Comparison of the Estimated Entity Trust Scores (the Case of a “Good” User)}
\label{fig:9} 
\end{figure}

\begin{figure}
\centering
\includegraphics*[width=0.5\textwidth]{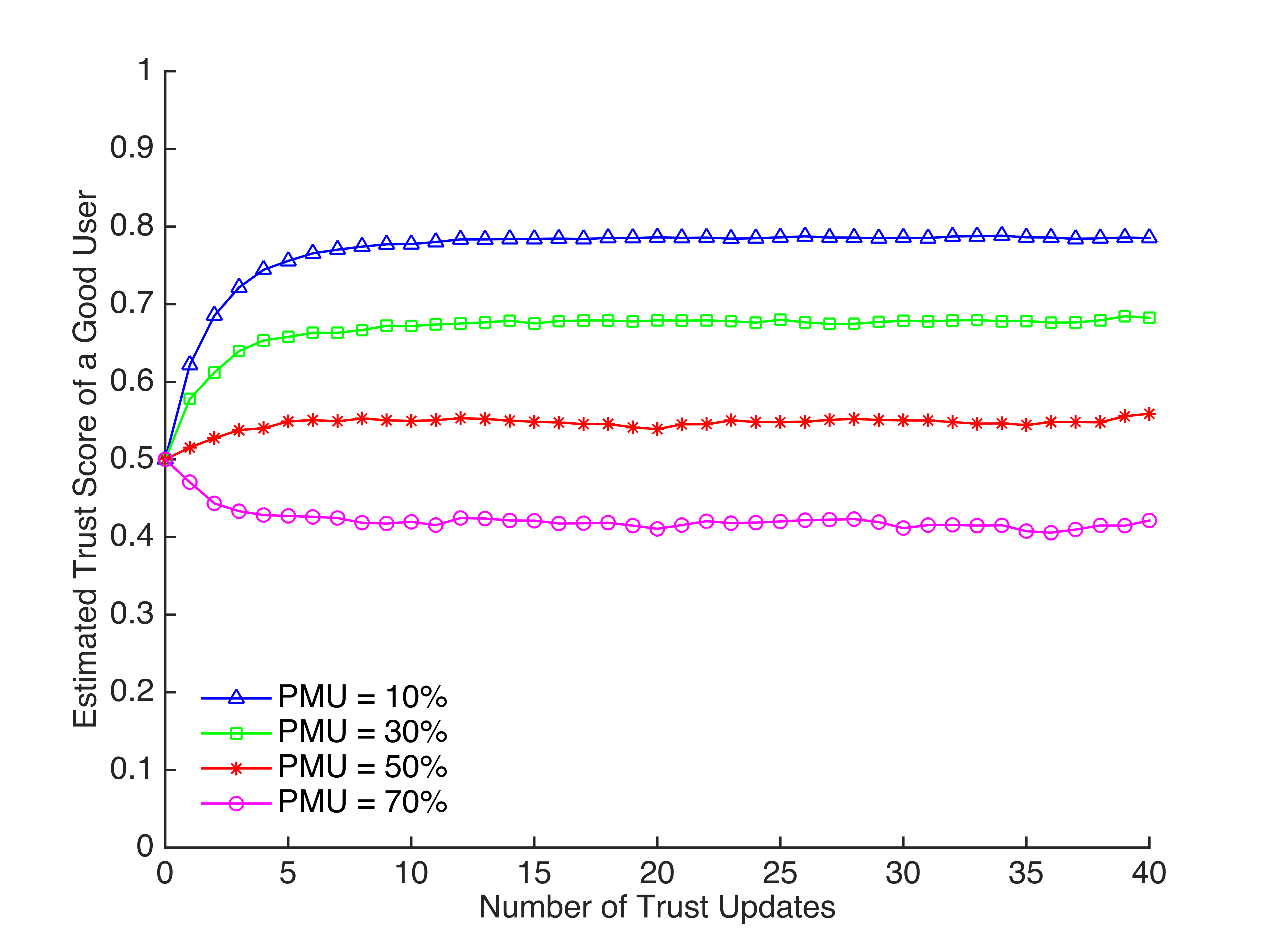}
\caption{Estimated Entity Trust Scores of a Good User Assessed by \cite{al2017trust}}
\label{fig:10} 
\end{figure}

\par Fig.~\ref{fig:9} compares the estimated entity trust score of a “good” user with the results obtained by implementing the scheme proposed in \cite{al2017trust}. The simulation setup for this experiment is in accordance with Table~\ref{fig:3}. As can be seen, the estimation accuracy is higher compared to \cite{al2017trust}. Also, as shown in Fig.~\ref{fig:10}, the trust value assessed by \cite{al2017trust} declines sharply as PMU increases. In fact, \cite{al2017trust} is very much susceptible to the higher presence of malicious users to the extent that it misidentifies good users as malicious for the PMU of 50\% and 70\%. In comparison, as previously shown in Fig.~\ref{fig:5} , when tuned with reasonable values for misdetection probabilities, our proposed scheme for entity trust estimation performs very well. The higher accuracy of our proposed scheme can be attributed to our Bayesian learning approach for entity trust estimation. The learning algorithm is very much effective in identifying the users’ inclinations towards contributing observations. The algorithm is also fairly robust against the increase of misdetection probabilities. On the contrary, the trivial majority-based judgment used in \cite{al2017trust} \linebreak quickly loses its efficiency when malicious behavior begins to dominate.

\par
\begin{figure}
\centering
\includegraphics*[width=0.5\textwidth]{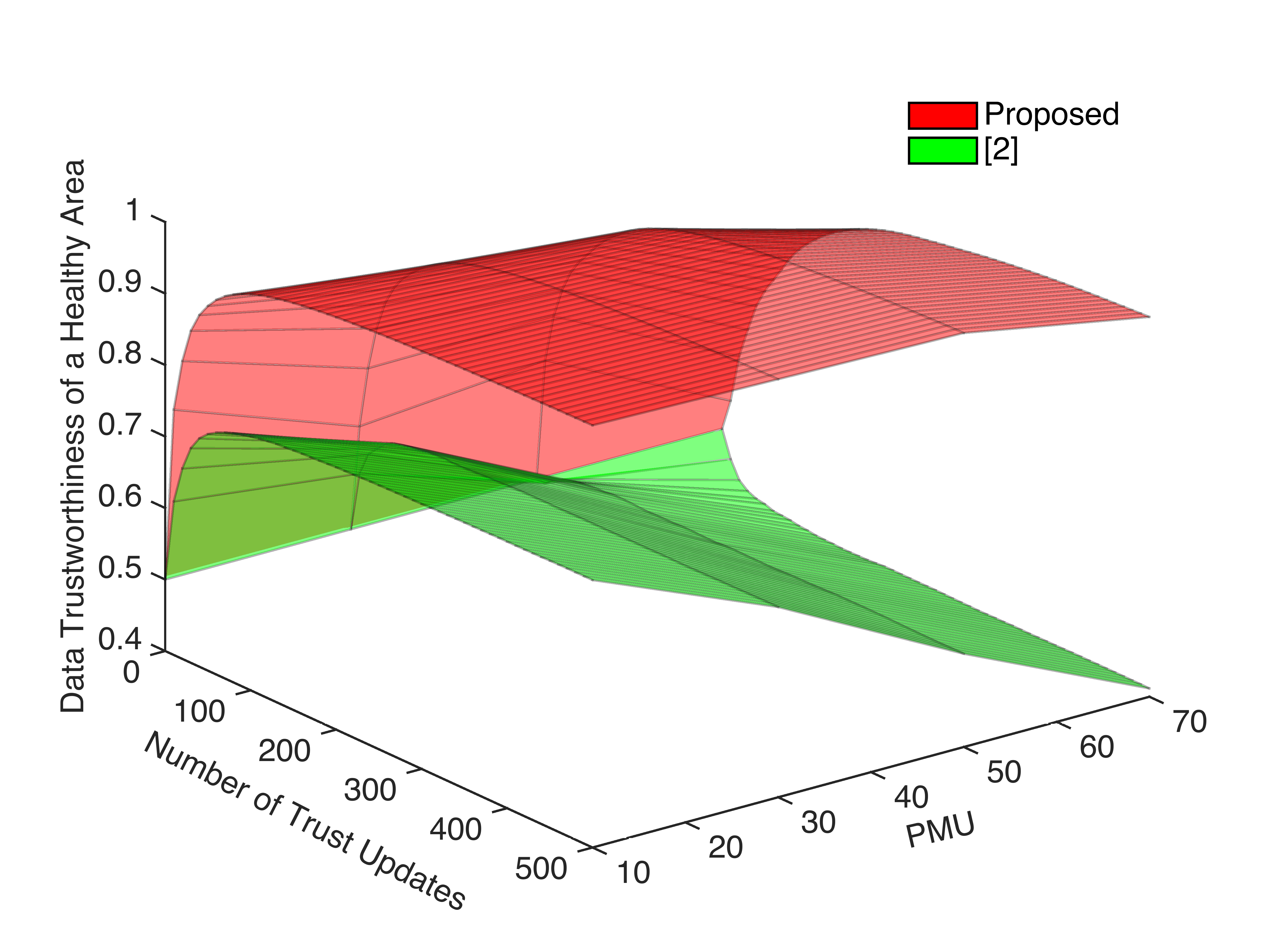}
\caption{Comparison in Terms of Data Trustworthiness (the case of a “Healthy” Area)}
\label{fig:11} 
\end{figure}

 Finally, Fig.~\ref{fig:11} shows the results of comparison in terms of the data trustworthiness for truly healthy areas. As a general observation, the accuracy drops as the value of PMU increases. However, in the presence of a high percentage of malicious users, the data trustworthiness in \cite{al2017trust} is much lower compared to the proposed method. The higher accuracy of the proposed scheme lies in the fact that it incorporates both the contextual parameters and the entity trust of the contributor for assessing the weight of observations. In contrast, apart from its lower quality estimates for entity trust scores, \cite{al2017trust} assigns weights to each observation merely based on the entity trust of the contributor and aggregates these weights using a simple averaging method.

\section{Conclusion}
\label{sec:5}
In this paper, we have proposed a hybrid entity/data trust management scheme for an IoT-enabled environmental \linebreak health/accessibility monitoring service. We envisaged a centralized service that collects observations from the users, and evaluates their contributions. In return, it also responds to user queries about the status of the geographic region under surveillance. We proposed a Bayesian learning-based procedure to estimate the trust score of the users (entities), which is an effective procedure in identifying the behavioral inclinations of the users towards reporting correct/false observations. As for data fusion, we leveraged on Dempster-Shafer Theory to fuse the observations, and to assess the trustworthiness of the data by using the freshly estimated users’ trust scores as well as the contextual properties associated with the observations. We conducted simulation experiments to evaluate the performance of our proposed trust management scheme. More specifically, we evaluated the accuracy of our scheme for both entity and data trust under different misdetection probabilities and against increasing population of malicious users. Moreover, we demonstrated the resiliency of our proposed scheme against the behavioral changes of the users. We also conducted a comparative evaluation against a closely-related prior work. The comparisons have demonstrated the superiority of our trust management scheme in terms of the accuracy of the estimated trust, and its robustness against higher presence of malicious users. In future, we plan to propose a trust management scheme that utilizes both social and QoS metrics to compute trust scores and will use both central and distributed propagation methods to estimate trust.


\section*{Appendix A: Proof of Theorem ~\ref{th:1}}
\textit{\textbf{Proof.}} Since a Markovian property holds between the following three events $S_n=s_j$, $b_{n,t}$ and $E_{n,t}$, the future state of each event is independent of the sequence preceded it and only depends on the present state of the event; i.e.,

\begin{equation*}
\begin{split}
\mathbb{P}\left[E_{n,t},b_{n,t}|S_n\right]=\mathbb{P}\left[E_{n,t}|b_{n,t},S_n\right]\times \mathbb{P}\left[b_{n,t}|S_n\right]= \\
\mathbb{P}\left[E_{n,t}|b_{n,t}\right]\times \mathbb{P}\left[b_{n,t}|S_n\right].
\end{split}
\end{equation*}

\par Given that the user’s behavior in contributing observations is either “correct” or “wrong”, we can derive that:

\begin{equation*}
\begin{split}
\mathbb{P}\left[E_{n,t}|S_n=s_j\right]=\ \mathbb{P}\left[E_{n,t},b_{n,t}\mathrm{=}\mathbb{C}\mathrm{|}S_n=s_j\right]+ \\
\mathbb{P}\left[E_{n,t},b_{n,t}=\mathbb{W}|S_n=s_j\right]=\mathbb{P}\left[E_{n,t}|b_{n,t}\mathrm{=}\mathbb{C}\right]s_j+ \\
\mathbb{P}\left[E_{n,t}|b_{n,t}=\mathbb{W}\right]{(1-s}_j).
\end{split}
\end{equation*}

Now, since $E^{l,k}_{n,t}={\{e^{l,k,i}_{n,t}\}}_{i=1:M^{l,k}_{n,t}}$ , and all evaluations $e^{l,k,i}_{n,t}$ are independent of each other, it follows that:

\begin{equation*}
\mathbb{P}\left[E_{n,t}|b_{n,t}\right]=\prod^{M_{n,t}}_{i=1}{\mathbb{P}\left[e^i_{n,t}|b_{n,t}\right]}.
\end{equation*}

\par Furthermore, if a user’s behavior is to contribute “correct” observations, it can be derived that:

\begin{equation*}
\begin{split}
\mathbb{P}\left[E_{n,t}|b_{n,t}\mathrm{=}\mathbb{C}\right]=\prod^{N_t{(}\mathbb{C}{)}}_{i=1}{\mathbb{P}\left[e^i_{n,t}=\mathbb{C}|b_{n,t}=\mathbb{C}\right]}{\times} \\
\prod^{N_t{(}\mathbb{W}{)}}_{i=1}{\mathbb{P}\left[e^i_{n,t}=\mathbb{W}|b_{n,t}=\mathbb{C}\right]}={{(f}_P)}^{N_t{(}\mathbb{W}{)}}{\times}{{(1-f}_P)}^{N_t{(}\mathbb{C}{)}}.
\end{split}
\end{equation*}

\par Similarly, if the user’s behavior is to contribute “wrong” observations, we have:

\begin{equation*}
\mathbb{P}\left[E_{n,t}|b_{n,t}\mathrm{=}\mathbb{W}\right]={{(f}_N)}^{N_t{(}\mathbb{C}{)}}{\times}{{(1-f}_N)}^{N_t{(}\mathbb{W}{)}}.
\end{equation*}

\par Hence,

\begin{equation*}
\begin{split}
{\mathbb{P}}\left[E_{n,t}|S_n=s_j\right]=s_j\times {\left(f_P\right)}^{{\mathcal{N}}_t(\mathbb{W})}\times {\left({1-f}_P\right)}^{{\mathcal{N}}_t\mathrm{(}\mathbb{C}\mathrm{)}}+ \\
\left(1-s_j\right)\times {\left(f_N\right)}^{{\mathcal{N}}_t\mathrm{(}\mathbb{C}\mathrm{)}}\times {\left({1-f}_N\right)}^{{\mathcal{N}}_t(\mathbb{W})}.
\end{split}
\end{equation*}

\bibliographystyle{spbasic}      
\bibliography{references.bib}   

\end{document}